\documentclass[11pt]{article}
\usepackage{amsmath}
\usepackage{ifthen}
\usepackage{setspace}

\newboolean{tab}
\setboolean{tab}{false}
\usepackage{graphics}

\usepackage[latin1]{inputenc}
 \usepackage[authoryear]{natbib}
\usepackage{graphicx}
\usepackage{latexsym}
\usepackage{amssymb}
\usepackage{epsfig}

\makeatletter\@addtoreset{equation}{section}\makeatother

\def\eps{\varepsilon}
\def \R{\mathbb{R}}
\def \E{\mathbb{E}}
\def \N{\mathbb{N}}

\def \Cov{\mbox{Cov}}

\newcommand{\ef}{\mathcal{F}}

\newcommand{\Dkonv}{\stackrel{\mathcal{D}}{\rightarrow}}
\newcommand{\Pkonv}{\stackrel{P}{\rightarrow}}

\newcommand{\Yc}{\mathcal{Y}}
\newcommand{\Wc}{\mathcal{W}}
\newcommand{\Xc}{\mathcal{X}}

\newcommand{\Kc}{\mathcal{K}}

\newcommand{\Fc}{\mathcal{F}}

\newcommand{\Nc}{\mathcal{N}}

\newcommand{\Tc}{\mathcal{T}}

\newcommand{\Zc}{\mathcal{Z}}

\newcommand{\Gb}{\mathbb{G}}

\newcommand{\bean}{\begin{eqnarray*}}
\newcommand{\eean}{\end{eqnarray*}}
\newcommand{\bea}{\begin{eqnarray}}
\newcommand{\eea}{\end{eqnarray}}
\newcommand{\be}{\begin{eqnarray}}
\newcommand{\ee}{\end{eqnarray}}
\newcommand{\beq}{\begin{equation}}
\newcommand{\eeq}{\end{equation}}

\newtheorem{theo}{Theorem}[section]
\newtheorem{lemma}[theo]{Lemma}

\newtheorem{rem}[theo]{Remark}

\newtheorem{example}[theo]{Example}

\begin{document}

\title{Smoothed quantile regression processes for binary response models.}

\author{Stanislav Volgushev\\
{\it University of Toronto} \thanks{The idea of considering binary response quantile processes originated from discussions with Prof. Roger Koenker. I am thankful to him for the encouragement and many insightful discussions on this topic. My thanks also go to Prof. Jiaying Gu for many helpful discussions. Any remaining mistakes are my sole responsibility. I also thank the Editor Prof. Peter C.B. Phillips, the co-Editor Prof. Yoon-Jae Whang and three anonymous Referees for constructive and insightful comments on previous versions of this manuscript that helped to considerably improve the presentation and content of this paper. Part of this research was conducted while I was a visiting scholar at UIUC. I am very grateful to the Statistics and Economics departments for their hospitality. Financial support from the DFG (grant VO1799/1-1) and from a discovery grant from NSERC of Canada is gratefully acknowledged.}}

\maketitle

\begin{abstract}
In this paper, we consider binary response models with linear quantile restrictions. Considerably generalizing previous research on this topic, our analysis focuses on an infinite collection of quantile estimators. We derive a uniform linearisation for the properly standardized empirical quantile \textit{process} and discover some surprising differences with 
the setting of continuously observed responses. Moreover, we show that considering quantile processes provides an effective way of estimating binary choice probabilities without restrictive assumptions on the form of the link function, heteroskedasticity or the need for high dimensional non-parametric smoothing necessary for approaches available so far. A uniform linear representation and results on asymptotic normality are provided, and the connection to rearrangements is discussed.
\end{abstract}
\section{Introduction}

In various situations in daily life, individuals are faced with making a decision that can be described by a binary variable. Examples relevant to various fields of economics include the decision to participate in the labour market, to retire, to make a major purchase. From an econometric point of view, such decisions can be modelled by a binary response variable $Y = I\{Y^* \geq 0\}$ that depends on an unobserved continuous random variable $Y^*$ which summarizes an individual's preferences. In the presence of covariates, say $W$, a natural question is: what can we infer about the distribution of the unobserved variable $Y^*$ conditional on $W$ from observations of i.i.d. replicates of $(Y,W)$? In a seminal paper, \cite{manski1975} assumed that $Y^* = W^T\beta + \eps$ where the `error' $\eps$ satisfies the conditional median restriction $P(\eps\leq 0|W=w) = 0.5$ and derived conditions on the distribution of $(\eps,W)$ that imply identifiability of the coefficient vector $\beta$ up to scale. In later work, \cite{manski1985} extended those results to general quantile restrictions of the form $P(\eps\leq 0|W=w) = \tau$ for fixed $\tau\in (0,1)$. A more detailed discussion of identification issues was provided in \cite{manski1988}. Due to their importance in understanding binary decisions, binary choice models have ever since aroused a lot of interest and many estimation procedures have been proposed [see \cite{cosslett1983}, \cite{horowitz1992}, \cite{PoStSt1989}, \cite{ichimura1993}, \cite{KlSp1993}, \cite{coppejans2001}, \cite{kordas2006} and \cite{khan2013} to name just a few]. 

A particularly challenging part of analysing binary response models lies in understanding the stochastic properties of corresponding estimation procedures. The asymptotic distribution of Manski's estimator was derived in \cite{KiPo1990} under fairly general conditions, while a non-standard case was considered in \cite{portnoy1998}. In particular, \cite{KiPo1990} demonstrated that the convergence rate is $n^{-1/3}$ and that the limiting distribution is non-Gaussian. A different approach based on non-parametric smoothing that avoids some of the difficulties encountered by Manski's estimator was taken by \cite{horowitz1992}. By smoothing the objective function, \cite{horowitz1992} obtained both - better rates of convergence and a normal limiting distribution. However, note that the smoothness conditions on the underlying model are stronger than those of \cite{KiPo1990}. 

The approaches of Manski and Horowitz have in common that only estimators for the coefficient vector $\beta$ are provided. While those coefficients are of interest and can provide valuable structural information, their interpretation can be quite difficult since the scale of $\beta$ is not identifiable from the observations. On the other hand, the `binary choice probabilities' $p_w := P(Y=1|W=w)$ provide a much simpler and more straightforward interpretation.

Most of the available methods for estimating binary choice probabilities are of two basic types. The first and more thoroughly studied approach is to assume a model of the form $Y^* = W^T\beta + \eps$ where the $\eps$ is assumed to be either independent of $W$ [see \cite{cosslett1983} and \cite{coppejans2001}], or admit a very special kind of heteroskedasticity [\cite{KlSp1993}]. Another popular approach has been to embed the problem into general estimation of single index models, see for example \cite{PoStSt1989} or \cite{ichimura1993}. Here, it is again necessary to assume independence between $\eps$ and the covariate $W$. 

While in the settings described above it is possible to obtain parametric rates of convergence for the coefficient vector $\beta$ and also construct estimators for choice probabilities, in many cases the assumptions on the underlying model structure seem too restrictive.

An alternative approach allowing for general forms of heteroskedasticity was recently investigated by \cite{khan2013}, who proved that under general smoothness conditions any binary response model with $Y^* = W^T\beta + \eps$ is observationally equivalent to a Probit/Logit model with multiplicative heteroskedasticity, that is a model where $\eps = \sigma_0(W)U$ with $U$ independent of $W$ and general scale function $\sigma_0$. \cite{khan2013} also proposed to simultaneously estimate $\beta$ and the function $\sigma_0$ by a semi-parametric sieve approach. The resulting model allows one to obtain an estimator of the binary choice probabilities.
While this idea is extremely interesting, it effectively requires estimation of a $d$-dimensional function in a non-parametric fashion. For the purpose of estimating $\beta$, the function $\sigma_0$ can be viewed as a nuisance parameter and its estimation does not have an impact on the rate at which $\beta$ is estimable. However, the binary choice probabilities explicitly depend on $\sigma_0$ and can thus only be estimated at the corresponding $d$-dimensional non-parametric rate. In settings where $d$ is moderately large this can be quite problematic.

In the classical setting where responses are observed completely, linear quantile regression models [see \cite{KoBa1978}] have proved useful in providing a model that can incorporate general forms of heteroskedasticity and at the same time avoid non-parametric smoothing. In particular, by looking at a collection of quantile coefficients indexed by the quantile level $\tau$ it is possible to obtain a broad picture of the conditional distribution of the response given the covariates. The aim of the present paper is to carry this approach into the setting of binary response models. In contrast to existing methods, we can allow for rather general forms of heteroskedasticity and at the same time estimate binary choice probabilities without the need of non-parametrically estimating a $d$-dimensional function.   

The ideas explored here are closely related to the work of \cite{kordas2006}. Yet, there are many important differences. First, in his theoretical investigations, \cite{kordas2006} considered only a finite collection of quantile levels. The present paper aims at considering the quantile \textit{process}. Contrary to the classical setting, and also contrary to the results suggested by the analysis in \cite{kordas2006}, we see that the asymptotic distribution is a white noise type process with limiting distributions corresponding to different quantile levels being independent. An intuitive explanation of this seemingly surprising fact along with rigorous theoretical results can be found in Section \ref{sec:coef}. We thus provide both a correction and considerable extension of the findings in \cite{kordas2006}.\\
Further, our results on the quantile process pave the way to obtaining an estimator for the conditional probabilities $p_w$ and derive its asymptotic representation. While a related idea was considered in \cite{kordas2006}, no theoretical justification of its validity was provided. Moreover, we are able to considerably relax the identifiability assumptions that were implicitly made there. Finally, we demonstrate that our ideas are closely related to the concept of rearrangement [see \cite{DeNePi2006} or \cite{ChFeGa2010}] and provide new theoretical insights regarding certain properties of the rearrangement map that seem to be of independent interest.\\
The rest of the paper is organized as follows. In Section \ref{sec:coef}, we formally state the model and provide results on uniform consistency and a uniform linearisation of the binary response quantile process. All results hold uniformly over an infinite collection of quantiles $T$. In Section \ref{sec:prob}, we show how the results from Section \ref{sec:coef} can be used to obtain estimators of choice probabilities. We elaborate on the connection of this approach to rearrangements. A uniform asymptotic representation for a properly rescaled version of the proposed estimators is provided and their joint asymptotic distribution is briefly discussed. Theoretical properties of the rearrangement operator are collected in Section~\ref{sec:rearr}. Section~\ref{sec:sim} contains comments on practical implementation of the proposed estimation procedures together with a small simulation study. All proofs are deferred to an appendix. 

\section{Estimating the coefficients}\label{sec:coef}

Before we proceed to state our results, let us briefly recall some basic facts about identification in binary response models and provide some intuition for the estimators of \cite{manski1975} and \cite{horowitz1992}. Throughout the paper, we consider the following data-generating process
\begin{itemize}
\item[(M)] Assume that we have n i.i.d. replicates, say $(Y_i, W_i)_{i=1,...,n}$, drawn from the distribution $(Y=I\{Y^* \geq 0\},W)$ with $Y^*$ denoting the unobserved variable of interest and $W$ denoting a vector of covariates. Further, denote by $q_\tau(w)$ the conditional quantile function of $Y^*$ given $W=w$. Assume that for $\tau \in T \subset (0,1)$ we have $q_\tau(w) = w^T\beta_\tau$ for some vectors $\beta_\tau \in \R^{d+1}$.
\end{itemize} 

Observing that $I\{Y^*\geq 0\} = I\{aY^*\geq 0\}$ with $a>0$ arbitrary directly shows that the scale of the vector $\beta_\tau$ can not be identified from $(Y,W)$. The vector $\beta_\tau$ is identified up to scale if for example $\gamma\neq\beta_\tau$ implies that the distribution of $Y$ conditional on $W^T\beta_\tau>0$ differs from that conditional on $W^T\gamma>0$ on a sufficiently large set. More precisely, assume that the function $u \mapsto F_{Y^*|W}(w^T\beta_\tau + u|w)$ is strictly increasing for $u$ in a neighbourhood of zero and all $w \in \mbox{support(W)}$. In that case we have by the definition of the $\tau$'th quantile
\[
P(Y=1|W=w)  
\left\{
\begin{array}{cc}
> 1-\tau, & \mbox{if} \ w^T\beta_\tau >0 \\
= 1-\tau, & \mbox{if} \ w^T\beta_\tau = 0 \\
< 1-\tau, & \mbox{if} \ w^T\beta_\tau < 0.
\end{array}
\right.
\]
This suggests that the expectation of $(Y-(1-\tau))$ conditional on $W=w$ is positive for $w^T\beta_\tau>0$ and negative for $w^T\beta_\tau<0$. We thus expect that under appropriate conditions the function 
\[
S_\tau(\beta) := \E\Big[(Y - (1-\tau))I\{W^T\beta \geq 0\}\Big]
\]
should be maximal at $a\beta_\tau$ for any $a>0$. Consider a vector $\gamma\in \R^{d+1}$. Then
\[
S_\tau(\gamma)-S_\tau(\beta_\tau) = D_1(\gamma,\tau) + D_2(\gamma,\tau) 
\]
with
\bean
D_1(\gamma,\tau) &:=& \E[(Y-(1-\tau))I\{W^T\gamma\geq 0>W^T\beta_\tau\}],
\\
D_2(\gamma,\tau) &:=&  - \E[(Y-(1-\tau))I\{W^T\gamma<0 \leq W^T\beta_\tau\}]. 
\eean
Note that both quantities are non-positive, and at least one of them being strictly negative is sufficient for inferring $\beta_\tau/\|\beta_\tau\| \neq \gamma/\|\gamma\|$ from the observable data  for a fixed quantile $\tau\in T$. An overview and more detailed discussion of related results is provided in Chapter 4 of \cite{horowitz2009}. Assumptions ensuring identifiability for a collection of quantiles $\tau \in T$ are discussed below. 

A common assumption [see e.g. Chapter 4 in \cite{horowitz2009}] is that one component of $\beta_\tau$ is either constant or at least bounded away from zero. Without loss of generality, we assume that this holds for the first component of $\beta_\tau$. In order to simplify the notation of what follows, write the covariate $W$ in the form $W^T = (Z,X^T)$ with $Z$ being the first component of $W$ and $X$ denoting the remaining components. Denote the supports of $X,Z,W$ by $\Xc,\Zc,\Wc$, respectively. Define the empirical counterpart of $S_\tau$ by 
\[
\tilde S_{n,\tau}(\beta) := \frac{1}{n}\sum_{i=1}^n (Y_i - (1-\tau))I\{W^T_i\beta \geq 0\}
\]
and consider a smoothed version 
\[
\hat S_{n,\tau}(\beta) := \frac{1}{n}\sum_{i=1}^n (Y_i - (1-\tau))\Kc\Big(\frac{W^T_i\beta}{h_n}\Big)
\]
with $h_n$ denoting a bandwidth parameter and $\Kc(u) := \int_{-\infty}^u K(v)dv$ a smoothed version of the indicator function $I\{u \geq 0\}$. 
Following \cite{horowitz2009}, define the estimator $(\hat s_\tau,\hat b_\tau)$ through
\beq\label{eq:esti}
(\hat s_\tau,\hat b_\tau) = \mbox{argmax}_{s=\pm 1,b\in B} \hat S_{n,\tau}((s,b^T)^T),
\eeq
where $B \subset \R^d$ is some fixed compact set not depending on $\tau$. Note that we do not impose $\hat s_\tau = \hat s_{\tau'}$ in the estimation. In principle, this would be possible, but it would require us to consider several minimizations simultaneously and result in computational difficulties. Uniform consistency of $\hat s_\tau$ proved below implies that all values of $\hat s_\tau$ for different $\tau \in T$ will be equal with probability tending to one.

\begin{rem}{\rm
The proofs of all subsequent results implicitly rely on the fact that we \textit{know} which coefficient stays away from zero and that the covariate corresponding to this particular coefficient has a `nice' distribution conditional on all other covariates [see assumptions (F1), (D2) etc.]. This is in line with the approach of \cite{horowitz1992} and \cite{kordas2006} and makes sense in many practical examples. Results similar to the ones presented below might continue to hold if we use Manski's normalization $\|\hat\beta\|=1$ instead of setting the `right' component to $\pm 1$. However, the asymptotic representation would be somewhat more complicated. For this reason, we leave this interesting question to future research.}  $\hfill \blacksquare$    
\end{rem}

\begin{rem}{\rm As pointed out by the Editor Prof. P.C.B. Phillips, neither the approach of \cite{horowitz1992} nor Manski's normalization $\|\beta_\tau\| = 1$ work when the sub-vector of $\beta_\tau$ which does not include the intercept is the zero vector. In particular, this is the case when the vector of covariates $W$ and the latent variable $Y^*$ are independent. The latter model is observationally equivalent to any other model where $Y$ is independent of $W$.} $\hfill \blacksquare$ 
\end{rem}

\begin{rem}{\rm
Note that, due to the scaling, the estimator $\hat b_\tau$ defined above is an estimator of the re-scaled quantity $\bar b_\tau :=  b_\tau/|\beta_{\tau,1}|$ where $b_\tau := (\beta_{\tau,2},...,\beta_{\tau,d+1} )^T$. When interpreting the estimator $\hat b_\tau$, this must be taken into account. In particular, $\hat b_\tau$ can not be interpreted as a classical quantile regression coefficient. The fact that we divide by $|\beta_{\tau,1}|$ also explains the reason behind assumption (A). Some comments on the practical choice of this normalization are given in Section~\ref{sec:sim}.} $\hfill \blacksquare$ 
\end{rem}

In all of the subsequent developments we make the following basic assumption.

\begin{enumerate}
\item[(A)] The coefficient $\beta_{\tau,1}$ satisfies $\inf_{\tau\in T}|\beta_{\tau,1}| > 0$ and the coefficient $\beta_{\tau,1}$ has the same sign for all $\tau\in T$. In what follows, denote this sign by $s_0$.
\end{enumerate}

In order to establish uniform consistency of the smoothed maximum score estimator, we need the following assumptions.

\begin{enumerate}
\item[(K1)] The function $\Kc$ is uniformly bounded and satisfies $\sup_{|v|\geq c}|\Kc(v)-I\{v\geq 0\}| \to 0$ as $c\to\infty$.
\item[(F1)] The conditional distribution function of $Z$ given $X$, say $F_{Z|X}$, is uniformly continuous uniformly over $x\in\Xc$, that is
\[
\sup_{x\in \Xc}\sup_{v\in \R} \sup_{|u|\leq\delta} |F_{Z|X}(v|x) - F_{Z|X}(v+u|x)| \to 0 \quad \mbox{as} \quad \delta\to 0 .
\]
\item[(D1)] For any fixed $\tau\in T$, $\bar \beta_\tau = (s_0,\bar b_\tau^T)^T$ is the unique minimizer of $S_\tau(\beta)$ on $\{-1,1\}\times B$ and additionally
\[
d(\eps) :=  \inf_{\tau\in T}\inf_{\|\beta-\bar \beta_\tau\|\geq \eps, |\beta_1|=1} |S_\tau(\beta)-S_\tau(\bar \beta_\tau)| > 0 \quad \forall \eps>0.
\]
\end{enumerate}

In order to intuitively understand the meaning of condition (D1) above, note that conditions (K1) and (F1) imply that $\hat S_{n,\tau}(\beta) \to S_\tau(\beta)$ uniformly in $\tau,\beta$. Condition (D1) essentially requires that the maximum of $S_\tau(\beta)$ is `well separated' uniformly in $\tau$, which allows to obtain uniform consistency of a sequence of maximizers of any function that uniformly converges to $S_\tau$. Next, we give a simple condition on densities and distributions which implies (D1). This condition is inspired by the approach taken in \cite{manski1985,horowitz1992}.

\begin{itemize}
\item[(S)] The support of the distribution of $X$ is not contained in any proper linear subspace of $\R^{d}$. Additionally, the conditional density $f_{Z|X}(\cdot|x)$ of $Z$ given $X=x$ exists for all $x\in \Xc$ and $\inf_{x\in \Xc}\inf_{z\in [-D,D]} f_{Z|X}(z|x) > 0$ for some $D> \sup_{b\in B, x\in \Xc}|b^T x|$.
\end{itemize}

The fact that (D1) follows from (A), (M) and (S) is proved in the appendix.

\begin{lemma}\label{lem:unifcons}
Under assumptions (M), (K1), (D1), (F1) let $h_n\to 0$. Then the estimator $(\hat s,\hat b_\tau)$ is weakly uniformly consistent, that is
\[
\sup_{\tau \in T} \|(\hat s_\tau,\hat b_\tau) - (s_0,\bar b_\tau)\| = o_P(1).
\] 
\end{lemma}

The next collection of assumptions is sufficient for deriving a uniform linearization for $\hat b_\tau.$ Assume that there exist $\eta>0, k_0 \geq 2$ such that the following conditions hold.

\begin{enumerate}
\item[(K2)] The function $\Kc$ is two times continuously differentiable and its second derivative is uniformly H\"older continuous of order $\gamma>0$, that is it satisfies
\[ 
\sup_{|x-y|\leq \delta} |\Kc''(x) - \Kc''(y)| \leq C_K|x-y|^\gamma. 
\]
Denote the derivative of $\Kc$ by $K$. Assume that $K,K'$ are uniformly bounded, that $\int |v^2K'(v)| dv<\infty, \int |K'(v)|^2dv<\infty$, and that additionally $\alpha_n := h_n^{-1}\int_{|vh_n|>\eta} |K'(v)|dv = o(1)$. 
\item[(K3)] Assume that $\int |v^{k_0}K(v)| <\infty$, $\int v^j K(v)dv = 0$ for $1\leq j < k_0$, $\int_{|vh_n|>\eta} |K(v)|dv = o(h_n^{k_0})$ and that $\sup_{|a|>x}|K(a)| = o(x^{-1})$ as $x\to\infty$.
\item[(B)] The bandwidth $h_n$ satisfies $h_n = o(1)$ and additionally $(nh_n^3)^{-1/2}(\log n)^2 = o(1)$.
\item[(D2)] The distribution of $X$ (denoted by $P_X$ from now on) has bounded support $\Xc$ and the matrix $E[WW^T]$ exists and is positive definite. For almost every $x \in \Xc$, the covariate $Z$ has a conditional density $f_{Z|X}(\cdot|x)$ and
\[
\inf_{\tau \in T} \inf_{x \in \Xc}f_{Z|X}(-s_0x^T\bar b_\tau|x) > 0.
\]
\item[(D3)] For any vector $b$ with $\|b-\bar b_\tau\|\leq \eta$ the two functions $u \mapsto f_{Z|X}(s_0(-x^Tb + u)|x)$ and $u \mapsto F_{Y^*|X,Z}(0|x,s_0(-x^Tb + u))$ are two times continuously differentiable at every $u$ with $|u|\leq\eta$ for almost every $x \in \Xc$ and the first and second derivatives are uniformly bounded [uniformly over $x\in\Xc$,$\tau\in T$]. 
\item[(D4)] The function $u \mapsto f_{Z|X}(s_0(-x^T\bar b_\tau+u)|x)$ is $k_0-1$ times continuously differentiable for every $x\in\Xc$ at every $\tau \in T$ and $u$ with $|u|\leq\eta$. All derivatives, say $\partial_u^k f_{Z|X}(s_0(-x^T\bar b_\tau+u)|x)$, are uniformly bounded and satisfy: for all $\eps>0$ there exists $\delta > 0$ such that 
\[
\sup_{|u| \leq \eta, |u'| \leq \eta, |u-u'|\leq \delta} \sup_{\tau\in T} \sup_{x \in \Xc} \Big|\partial_u^k f_{Z|X}(s_0(-x^T\bar b_\tau+u)|x) - \partial_u^k f_{Z|X}(s_0(-x^T\bar b_\tau+u')|x)\Big| \leq \eps.
\]
The function  $u \mapsto F_{Y^*|X,Z}(0|x,s_0(-x^T\bar b_\tau + u))$ is $k_0$ times continuously differentiable  at every $\tau \in T$ and $u$ with $|u|\leq\eta$ at almost every $x\in \Xc$ and all derivatives, say $\partial_u^k F_{Y^*|X,Z}(0|x,s_0(-x^T\bar b_\tau + u))$, are uniformly bounded and satisfy: for all $\eps>0$ there exists $\delta > 0$ such that 
\[
\sup_{|u| \leq \eta, |u'| \leq \eta, |u-u'|\leq \delta} \sup_{\tau\in T} \sup_{x \in \Xc} \Big|\partial_u^k F_{Y^*|X,Z}(0|x,s_0(-x^T\bar b_\tau + u)) - \partial_u^k F_{Y^*|X,Z}(0|x,s_0(-x^T\bar b_\tau + u'))\Big| \leq \eps.
\]
\item[(D5)] The map $\tau\mapsto\beta_\tau$ is H\"older continuous of order $\gamma>0$ uniformly on $T$, that is we have $\sup_{\tau,\tau'\in T,|\tau-\tau'|\leq \delta} \|\beta_\tau - \beta_{\tau'}\| \leq C\delta^\gamma$ for some constant $C<\infty$ and all $\delta >0$.
\item[(Q)] For every $w\in \Wc$, the random variable $Y^*$ has a conditional density $f_{Y^*|W}$ conditional on $W = w$. Moreover,
\[
\inf_{\tau \in T} \inf_{x \in \Xc} f_{Y^*|X,Z}(0|x,-s_0x^T\bar b_\tau) > 0, \quad \sup_{\tau \in T} \sup_{x \in \Xc} f_{Y^*|X,Z}(0|x,-s_0x^T\bar b_\tau) < \infty.
\] 
\item[(D6)] The function $(y,w) \mapsto f_{Y^*|W}(y|w)$ is uniformly bounded on $\R \times \Wc$, i.e. 
\[
\sup_{y \in \R, w\in \Wc} f_{Y^*|W}(y|w) < \infty,
\]
and uniformly continuous on the set $\{|y|\leq \eta\}\times \Wc$. 
\end{enumerate}

The conditions on the kernel function $K$ are standard in the binary response setting and were for example considered in \cite{horowitz2009} and \cite{kordas2006}. Assumptions (D2)-(D4) are uniform versions of the conditions in \cite{horowitz1992} and are used to obtain results holding uniformly in an infinite collection of quantiles. Condition (D5) is used to obtain a rate in the uniform representation below. In particular, (D5) will follow with $\gamma = 1$ if there exists a set $\Wc_0 \subset \Wc$ such that the matrix $\E[WW^TI_{\{W\in \Wc_0\}}]$ is positive definite and $\inf_{w\in \Wc_0, \tau \in T} f_{Y^*|W}(w^T\beta_\tau|w) > 0$. Conditions (D6) and (Q) place mild restrictions on the conditional density of $Y^*$ given $W=w$. This kind of assumption is standard in classical quantile regression.

\begin{theo}\label{theo:betahat}
Under assumptions (A), (B), (M), (D1)-(D5), (F1), (K1)-(K3), (Q) we have
\beq \label{eq:bahres}
Q_0(\tau)(\hat b_\tau - \bar b_\tau) = - \tilde T_n(s_0,\bar b_\tau,\tau) + R_n(\tau),
\eeq
where
\bean
Q_0(\tau) &:=& |\beta_{\tau,1}|\int f_{Z|X}(-s_0x^T\bar b_\tau|x)f_{Y^*|X,Z}(0|x,-s_0x^T\bar b_\tau) xx^T dP_X(x),
\\
\tilde T_n(s,b,\tau) &:=& \frac{\partial \tilde S_{n,\tau}((s,b^T)^T)}{\partial b} = \frac{1}{nh_n}\sum_{i=1}^n (Y_i - (1-\tau))X_i K\Big(\frac{X_i^Tb + sZ_i}{h_n} \Big),
\\
\sup_{\tau\in T} \|R_n(\tau)\| &=& O_P(\kappa_n) := O_P\Big((h_n^{k_0} + (nh_n)^{-1/2}\log n)((nh_n^3)^{-1/2}\log n+h_n+\alpha_n)\Big),
\eean
and $\alpha_n$ is defined in assumption (K2). In particular, $\kappa_n  = o(h_n^{k_0} + (nh_n)^{-1/2})$ and thus $R_n$ is negligible compared to $\tilde T_n(s_0,\bar b_\tau,\tau)$.\\
Now assume that additionally condition (D6) holds and that $h_n^{2{k_0}} = o((nh_n)^{-1/2})$. Then, for any finite collection $\tau_1,...,\tau_K \in T$, we obtain
\beq \label{eq:betaweak}
\Big(\sqrt{nh_n}\Big(\hat b_{\tau_j} - \bar b_{\tau_j} - T_n(s_0,\bar b_{\tau_j},\tau_j)\Big)\Big)_{j=1,...,K} \Dkonv \Big((Q_0(\tau_j))^{-1}M_{\tau_j}\Big)_{j=1,...,K},
\eeq
where $\Dkonv$ denotes convergence in distribution,
\bean
T_n(s,b,\tau) &:=& \frac{h_n^{k_0}}{{k_0}!} \int v^{k_0} K(v)dv \int g_k(s,b,x) x dP_X(x), 
\\
g_j(s,b,x) &:=& \frac{\partial^j}{\partial u^j}\Big((\tau - F_{Y^*|X,Z}(0|x,s(-x^Tb+u)))f_{Z|X}(s(-x^Tb+u)|x) \Big)\Big|_{u=0},
\eean
$M_{\tau_i},M_{\tau_j}$ are independent for $j\neq i$ and 
\[
M_\tau \sim \Nc(0,\Sigma_\tau) 
\]
where
\[
\Sigma_\tau := \tau(1-\tau)\int K^2(u)du \int xx^T f_{Z|X}(-s_0x^T\bar b_\tau|x)dP_X(x).
\]
\end{theo}

Compared to the results available in the literature [e.g. in \cite{kordas2006} and \cite{horowitz2009}], the preceding theorem provides two important new insights. To the best of our knowledge, it is the first time that the estimator is simultaneously considered at an infinite collection of quantiles. Equally importantly, it demonstrates that the joint asymptotic distribution of several quantiles differs substantially from what both intuition and results in \cite{kordas2006} seem to suggest.

\begin{rem} \label{rem:joint}{\rm
In contrast to the setting when the response $Y^*$ is directly observed, the properly normalized quantile process at different quantile levels converges to independent random variables. An intuitive explanation for this surprising fact can be obtained from the asymptotic linearization in (\ref{eq:bahres}). For simplicity, assume that the kernel $K$ has compact support, say $[-1,1]$. To bound the covariance between the leading terms in the linearization of $\hat b_{\tau_1}, \hat b_{\tau_2}$, begin by observing that under the assumptions of Theorem~\ref{theo:betahat} we have for any $1\leq j,j'\leq d$ [see equation \eqref{eq:boundcov} in the Appendix for a proof]
\bean
&& \Big|\Cov\Big(\sqrt{nh_n}(\tilde T_n(s_0,\bar b_{\tau_1},{\tau_1}))_j, \sqrt{nh_n}(\tilde T_n(s_0,\bar b_{\tau_2},{\tau_2}))_{j'}\Big)\Big|
\\
&\leq& \frac{C}{h_n}\int\int \Big|K\Big(\frac{x^T \bar b_{\tau_1} + s_0z}{h_n} \Big)K\Big(\frac{x^T\bar b_{\tau_2} + s_0z}{h_n} \Big)\Big|f_{Z|X}(z|x)dzdP_X(x)
 + O(h_n).
\eean
Noting that $x^T\bar b_{\tau} + s_0z = w^T \beta_\tau/|\beta_{\tau,1}|$ with $w=(z,x)$, we find that $K\Big(\frac{x^T\bar b_{\tau_1} + s_0z}{h_n} \Big) K\Big(\frac{x^T\bar b_{\tau_2} + s_0z}{h_n} \Big) \neq 0$ if and only if $|w^T \beta_{\tau_k}| \leq h_n|\beta_{\tau_k,1}|, k = 1,2$, which implies $|w^T \beta_{\tau_1} - w^T \beta_{\tau_2}| \leq h_n (|\beta_{\tau_1,1}|+|\beta_{\tau_2,1}|)$. Next, note that $w^T \beta_\tau = F_{Y^*|W}^{-1}(\tau|w)$. Under assumption (D6) it follows that 
\[
|F_{Y^*|W}(y_1|w) - F_{Y^*|W}(y_2|w)| \leq |y_1-y_2| \sup_{(y,w)\in \R\times \mathcal{W}} f_{Y^*|W}(y|w) =: f_\infty |y_1-y_2|
\]
for any $w\in \Wc$. Substituting $y_k = F_{Y^*|W}^{-1}(\tau|w) = w^T\beta_{\tau_k}, k = 1,2$ we find that $|w^T(\beta_{\tau_1}-\beta_{\tau_2})| \geq f_\infty^{-1} |{\tau_1}-{\tau_2}|$ for any $w\in \Wc$. Hence for $\tau_1 \neq \tau_2$ the set $\{w\in \Wc: |w^T \beta_{\tau_k}| \leq h_n|\beta_{\tau_k,1}| \mbox{ for } k = 1,2\}$ will be empty for $h_n$ sufficiently small and asymptotic independence follows since the integral in the inequality above is identically zero as soon as $h_n$ is sufficiently small.

Intuitively speaking, all observations that have a non-zero contribution to $\tilde T_n(s_0,\bar b_\tau,\tau)$ will need to satisfy $|W_i^T\beta_\tau| \leq h_n |\beta_{\tau,1}|$. In particular, letting $h_n \to 0$ implies that asymptotically, for different values of $\tau$, disjoint sets of observations will be driving the distribution of $T_n$. Similar phenomena can be observed in other settings that include non-parametric smoothing, a classical example being density estimation.

Note that regarding this particular point the paper of \cite{kordas2006} contained a mistake. More precisely, \cite{kordas2006} claimed that the asymptotic distributions corresponding to different quantiles have a non-trivial covariance which is not the case. } $\hfill \blacksquare$ 
\end{rem}

\begin{rem}{\rm
As pointed out by a referee, asymptotic independence at different quantile levels does not occur in the setting of kernel-smoothing based quantile regression--see for instance \cite{chaudhuri1991}. This is due to the fact that the leading term in the Bahadur representation for kernel-based quantile regression typically is of the form $n^{-1}M_n(\tau,x) \sum_i W_n(X_i,x) (I\{Y_i \leq q_\tau\} - \tau)$. The weights $W_n(X_i,x)$ can often be bounded by $C_n I\{|X_i-x| \leq h_n\}$ for some constants $C_n$ and a bandwidth parameter $h_n \to 0$. However, the weights do not depend on $\tau$ so that for the same covariates but different quantiles the leading term is determined by the same set of observations. In contrast to that, the leading term in the present setting is $-(Q_0(\tau))^{-1}\frac{1}{nh_n}\sum_{i=1}^n (Y_i - (1-\tau))X_i K\Big(\frac{X_i^T\bar b_\tau + s_0Z_i}{h_n} \Big)$. Here, the quantile index $\tau$ appears in the kernel $K\Big(\frac{X_i^T\bar b_\tau + sZ_i}{h_n} \Big)$, and thus the quantile index also determines the set of observations which contribute to the asymptotic distribution of $\hat \beta_\tau$. This makes estimators at different quantiles asymptotically independent. 
 } $\hfill \blacksquare$ 
\end{rem}

The findings in Theorem \ref{theo:betahat} imply that there can be no weak convergence of the normalized process $\Big(\sqrt{nh_n}\Big(\hat b_{\tau} - \bar b_{\tau} - T_n(s_0,\bar b_{\tau},\tau)\Big)\Big)_{\tau\in T}$ in a reasonable functional sense since the candidate `limiting process' has a `white noise' structure and is not tight. This will present an additional challenge for the analysis of estimators for binary choice probabilities constructed in Section \ref{sec:prob}.  

Before proceeding to the estimation of conditional (choice) probabilities, we briefly comment on potential extensions of the findings in this section to parametric conditional quantile models that are not necessarily linear. 

\begin{rem}{\rm
As pointed out by the associate editor and a referee, it is natural to wonder if the methods discussed above can be extended to parametric non-linear models. The ideas discussed in this section can be generalized to settings where the latent variables $Y^*$ have conditional quantile functions of the form
\beq \label{eq:nonlin1}
q_\tau(x,z) = s_0 z + h(x,b_\tau), \quad \tau \in T
\eeq
for a set $T \subset (0,1)$, a parametric function $h$ that depends on the parameter $b$ which can vary with $\tau$ and where $s_0 \in \{-1,1\}$ is independent of $\tau$. Note that any model where the conditional quantile functions of the latent variable $Y^*$ are of the form 
\beq \label{eq:nonlin2}
q_\tau(x,z) = \beta_{\tau,1} z + \tilde h(x,\gamma_\tau), \quad \tau \in T
\eeq
with $\beta_{\tau,1}$ bounded away from zero uniformly on $T$ is observationally equivalent to a model where 
$q_\tau(x,z) = sgn(\beta_{\tau,1}) z + \tilde h(x,\gamma_\tau)/|\beta_{\tau,1}|$ [with $sgn(a)$ denoting the sign of $a$]. Thus the model specification in \eqref{eq:nonlin1} incorporates all models of the form \eqref{eq:nonlin2}, after re-parametrization. A generalization to other parametric models that do not have this simple additive structure seems possible, but developing the corresponding asymptotic results would require an approach that is different from the one in the present paper.

Estimators of the parameters in model \eqref{eq:nonlin1} can now be constructed in a similar fashion as in the linear case. More precisely, it seems natural to define
\beq \label{est:nonlin}
(\hat s_\tau,\hat b_\tau) = \mbox{argmax}_{s=\pm 1,b\in B} \hat S_{n,\tau}(s,b)
\eeq
where $B \subset \R^p$ is a compact set and
\[
\hat S_{n,\tau}(s,b) := \frac{1}{n}\sum_{i=1}^n (Y_i - (1-\tau))\Kc\Big(\frac{s Z_i + h(X_i,b))}{h_n}\Big).
\]
Under appropriate conditions on the parametrization and conditional distributions, the consistency results from Lemma~\ref{lem:unifcons} and asymptotic linearization from Theorem \ref{theo:betahat} can be generalized to the estimation of non-linear coefficients. Details are omitted for the sake of brevity.
$\hfill \blacksquare$
}\end{rem}

\section{Estimating conditional probabilities}\label{sec:prob}

Partly due to the lack of complete identification, the coefficients estimated in the preceding section might be hard to interpret. A more tractable quantity is given by the conditional probability $p_w := P(Y = 1|W=w)$. One possible way to estimate this probability is local averaging. However, due to the curse of dimensionality, this becomes impractical if the dimension of $W$ exceeds 2 or 3. An alternative is to assume that the linear model $q_\tau(w) = w^T\beta_\tau$ holds for all $\tau \in T \subset (0,1)$. By definition of $Y = I\{Y^*\geq 0\}$, the existence of $\tau_w \in T$ with $w^T\beta_{\tau_w} = 0$ implies that $p_w = 1-\tau_w$. Since the quantile function of $Y^*$ is given by $w^T\beta_\tau$ we obtain $P(Y^*\leq 0|W=w) = w^T\beta_{\tau_w}$. By definition of the quantile function and the assumptions on $Y^*$, $\tau<\tau_w \Leftrightarrow w^T\beta_{\tau}<w^T\beta_{\tau_w}$. This implies the equality $\tau_w = \int_0^1 I\{w^T\beta_\tau \leq 0 \} d\tau$. In particular, we have for any $(a,b)\subset T$ with $a<\tau_w<b$ 
\[
p_w = 1 - \tau_w = \int_0^1 I\{w^T\beta_\tau \geq 0\} d\tau = 1-b + \int_{a}^{b} I\{w^T\beta_\tau \geq 0\} d\tau = 1-b + \int_{a}^{b} I\{w^T\bar\beta_\tau \geq 0\} d\tau,
\] 
where the last equality follows since $w^T\beta_\tau = |\beta_{\tau,1}|w^T\bar\beta_\tau$ for all $w$ and $\tau$. This suggests to estimate $p_w$ by replacing $\beta_\tau$ in the above representation with the estimator $\hat\beta_\tau$ from the preceding section after choosing $(a,b)$ in some sensible manner. A similar representation was considered by \cite{kordas2006} with $a=0,b=1$. The fact that $\hat\beta$ is an estimator of the re-scaled version $\bar\beta_\tau$ is not important here since multiplication by a positive number does not affect the inequality $w^T\beta_\tau \geq 0$. From here on, define
\beq\label{hatpw}
\hat p_w(a,b) := 1-b + \int_{a}^{b} I\{w^T\hat\beta_\tau \geq 0\} d\tau.
\eeq

From the definition of $\hat p_w(a,b)$, it is not difficult to show [using Lemma \ref{lem:unifcons}] that under (A), (K1), (F1) and (D1), (Q) holding for $T = [a,b]$ we have
\begin{equation}\label{conspw}
\hat p_w(a,b) \Pkonv \left\{
\begin{array}{cl}
1 - a, & \mbox{if} \ p_w \geq 1 - a, \\
p_w, & \mbox{if} \ p_w \in (1-b,1-a), \\
1 - b, & \mbox{if} \ p_w \leq 1 - b.
\end{array}
\right.
\end{equation}

Under the same assumptions, the definition of $\hat p_w(a,b)$ implies that, with probability tending to one, $\hat p_w(a,b) = \hat p_w(a',b')$ as long as $a<\tau_w<b$ and $a'<\tau_w<b'$. This suggests that, from an asymptotic point of view, the choice of $a,b$ in the estimator $\hat p_w(a,b)$ is not very critical. However, $\tau_w$ is unknown in practice and some care must be taken in applications. Some comments on the practical choice of $a,b$ can be found in Section~\ref{sec:sim}.

\begin{rem}\label{rem:ab}{\rm
The representation in \eqref{hatpw} indicates that in order to estimate $p_w$ we do not need the linear model $q_\tau(w) = w^T\beta_\tau$ to hold globally and also do not require that $\beta_\tau$ can be estimated for all $\tau\in [0,1]$. In fact, the validity of the linear model $q_\tau(w) = w^T\beta_\tau$ for $\tau$ in a neighbourhood of $\tau_w$ and estimability of $\beta_\tau$ on this region is sufficient for the asymptotic developments provided below. This insight is interesting from a theoretical point of view. Making use of it in practice seems more difficult since prior knowledge on bounds $\tau_w$ for a given point $w$ might not be available. $\hfill \blacksquare$ 
}
\end{rem}

For values of $w$ for which there exists $\tau_w \in (a,b)$ with $w^T \beta_{\tau_w} = 0$, precise statements about the asymptotic distribution of $\hat p_w(a,b)$ are possible. To derive this distribution, we begin by observing that the definition of $\hat p_w$ is closely connected to the concept of rearrangement [see \cite{HaLiPo1988}]. More precisely,
recall that the monotone rearrangement $\Phi$ of a function $g:[0,1]\to\R$ is defined as 
\beq \label{eq:psi}
\Phi_g(u) = \Psi_g^{-}(u), \quad \Psi_g(v) = \int_0^1 I\{g(u)\leq v\}du
\eeq
where $\Psi_g^{-}$ denotes the generalized inverse of the function $u\mapsto \Psi_g(u)$. The first step of the rearrangement, $\Psi_g$, is the \textit{distribution function of $g$ with respect to Lebesgue measure}. Thus we can interpret the integral $\int_0^1 I\{w^T\beta_\tau \leq 0 \} d\tau$ in the definition of $\tau_w$ as the distribution function of the map $\tau \mapsto w^T\beta_\tau$. Previously, a smoothed version of the first step of the rearrangement was used by \cite{DeVo2008} to invert a non-increasing estimator of an increasing function in the setting of quantile regression. Properties of the rearrangement viewed as mapping between function spaces were considered, among others, in \cite{DeNePi2006} for estimating a monotone function and \cite{ChFeGa2010} for monotonising crossing quantile curves. However, existing results in the literature are not applicable in the present setting -- see Remark~\ref{rem:rearr} for a more detailed discussion. To keep the presentation simple, a more detailed discussion of the map $\Psi_g$ is deferred to Section~\ref{sec:rearr}, while the remaining part of the present section will be devoted to a closer study of the estimator $\hat p_w(a,b)$. \\

Before proceeding with the theory, we briefly comment on the practical implementation of the proposed estimators. In order to compute the integral in \eqref{hatpw}, we recommend to use an approximation of $\hat \beta_\tau$ on a uniformly spaced grid of the form $a = \tau_0<\tau_1<...<\tau_G=b$. If we choose a grid of width $o((nh_n)^{-1/2})$, Lemma \ref{lem:betaequi} in the appendix will ensure that the discrete approximation to $\hat \beta_\tau$ has the same first order asymptotic properties as the original estimator. As already discussed previously, we would recommend to choose $a,b$ as small and large as the data permit, respectively. Computing the estimators in \eqref{eq:esti} is a challenging problem since the optimization is not convex and the criterion function is likely to have multiple maxima. \cite{kordas2006} proposed to compute such estimators by means of the simulated annealing algorithm which was discussed in \cite{GoFeDe1994}. In general, finding reliable algorithms that allow to minimize non-convex functions is a very challenging practical problem. Since the focus of the present paper is more theoretical, we leave an answer to this very interesting question in the setting of binary response models to future research. For the unsmoothed version of the score function, a very interesting alternative which is based on Mixed Integer Programming was proposed by \cite{FlKo2008}. Those authors show that their approach can outperform existing methods by a large margin. However, they make use of the specific form of the unsmoothed maximum score estimator, and it is not clear if their approach can be utilized in the smoothed setting considered here.    

We now state the additional assumptions that will used to derive the limiting distribution of $\hat p_w$. Assume that for some $\delta>0$ the conditions of Theorem \ref{theo:betahat} hold on the set $T^\delta := [t_L-\delta,t_U+\delta]$ with $T := [t_L,t_U] \subset (\delta,1-\delta)$. This particular form of the set $T$ is assumed in order to simplify the definition of the estimator in \eqref{hatpw}. Consider the following conditions.

\begin{enumerate}
\item[(T)] The function $\tau \mapsto \beta_\tau$ is continuously differentiable on $T^\delta$ and its derivative is uniformly H\"older continuous of some order $\gamma' > 0$. The function $y \mapsto F_{Y^*|W}(y|w)$ is continuous for all $w \in \Wc$.  
\item[(K4)] The function $x\mapsto \mathcal{K}(x)$ is two times continuously differentiable. Write $K = \mathcal{K}'$ and assume that there exist $c_0 < \infty, \eps>0$ such that $K' = \mathcal{K}''$ satisfies $\sup_{|x|\geq c}|K'(x)|\leq c^{-1/2-\eps}$ for all $c\geq c_0$.
\end{enumerate}

Next, define the set 
\[
\Wc_T := \Big\{w\in\Wc \Big|~\exists \tau_w\in T = [t_L,t_U]: w^T\beta_{\tau_w} = 0 \Big\} = \Big\{w\in\Wc \Big|~ p_w \in [1-t_U, 1-t_L] \Big\}.
\] 
Note that under (T) the function $\tau \mapsto w^T \beta_\tau = F_{Y^*|W}^{-1}(\tau|w)$ is strictly increasing on $T^\delta$ for all $w \in \Wc_T$, so that the points $\tau_w$ in the definition of $\Wc_T$ are unique.

\begin{theo}\label{theo:pw}
Assume that for some $\delta>0$ the conditions of Theorem \ref{theo:betahat} hold on the set $T^\delta := [t_L-\delta,t_U+\delta]$ with $T := [t_L,t_U] \subset (\delta,1-\delta)$ and let conditions (K4) and (T) hold. Assume that for each $w\in \Wc_T$ we have $\tau_w\in(a_w,b_w) \subset T$ and that $h_n^{k_0} = O((nh_n)^{-1/2})$. Then for any $w\in \Wc_T$
\beq \label{eq:bahpw}
\hat p_w(a_w,b_w) - p_w = -w^T(\hat \beta_{\tau_w} - \bar \beta_{\tau_w})|\beta_{\tau,1}|f_{Y^*|W}(0|w) + R_n^{(2)}(w).
\eeq
Moreover, for any compact $\Wc_0 \subset \Wc_T$ 
\[
\sup_{w\in \Wc_0} |R_n^{(2)}(w)| = o_P((nh_n)^{-1/2}).
\] 
Finally, for any finite collection $w_1,...,w_k \in \Wc_T$ with $w_j^T = (z_j,x_j^T)$ we obtain
\bean
&& \Big(\sqrt{nh_n}\Big(\hat p_{w_j} - p_{w_j} + |\beta_{\tau,1}|f_{Y^*|W}(0|w_j)x_j^TT_n(s_0,\bar b_{\tau_{w_j}},\tau_{w_j})\Big)\Big)_{j=1,...,k} 
\\
&& \quad\quad\quad\quad \Dkonv \Big(|\beta_{\tau,1}|f_{Y^*|W}(0|w_j)x_j^T(Q_0(\tau_{w_j}))^{-1}M_{\tau_{w_j}}\Big)_{j=1,...,k}
\eean
where $T_n,M_\tau,Q_0$ are defined in Theorem \ref{theo:betahat}.
\end{theo}

\begin{rem}{\rm
As pointed out by a referee, the set $\Wc_T$ above Theorem~\ref{theo:pw} depends on the unknown $\tau_w$. In order to get a sense of whether the representation in~\eqref{eq:bahpw} holds for a given value $w$, observe that $\Wc_T = \{w\in\Wc |~ p_w \in [1-t_U, 1-t_L] \}$ and that $\hat p_w$ has the additional property that $\hat p_w(a,b) \in [1-b, 1-a]$ a.s. Hence the representation in~\eqref{eq:bahpw} is likely to hold if $\hat p_w(a_w,b_w)$ is not close to the boundary of $[1-b_w, 1-a_w]$ and can not be expected to be true when $\hat p_w(a_w,b_w)$ equals $1-a_w$ or $1-b_w$.
}$\hfill \blacksquare$ 
\end{rem}

From the results derived above, we see that the convergence rate of the estimators for binary choice probabilities corresponds to the rate typically encountered if one-dimensional smoothing is performed. Compared to the results of \cite{khan2013}, whose rates correspond to $d-$dimensional smoothing, this can be a very substantial improvement. While our assumptions are of course more restrictive than those of \cite{khan2013}, the form of allowed heteroskedasticity is more general than the simple multiplicative heteroskedasticity or even homoskedasticity assumed in previous work by \cite{cosslett1983}, \cite{KlSp1993}, \cite{coppejans2001}. While we of course do not suggest to completely replace the methodologies developed in the literature, we feel that our approach can be considered as a good compromise between flexibility of the underlying model and convergence rates. It thus provides a valuable supplement and extension of available procedures.      

\subsection{Properties of distribution functions with respect to Lebesgue measure} \label{sec:rearr}

In this section we state a general result that allows to derive a uniform linearization of the map $\Psi$ defined in \eqref{eq:psi}. In situations where a functional central limit result does not hold [this will often be the case in the situation of estimators built from local windows], this result is of independent interest. In particular, it can be used to derive a uniform Bahadur representation for the estimator $\hat p_w$ in the previous section.  

\begin{theo}\label{theo:rear}
Consider a collection of non-decreasing functions $g_{q}:[0,1]\to\R$ indexed by a general set $Q$ and assume that for all $q\in Q$ there exists $u_{0,q}\in(0,1)$ with $g_q(u_{0,q})=0$. Additionally, assume that there exists $\delta>0$ such that each $g_q$ is continuously differentiable in a neighbourhood $U_\delta(u_{0,q})\subset (0,1)\forall q\in Q$ and that $\inf_{q\in Q} g_q'(u_{0,q}) = g_{min} > 0$. 

Let $\chi(\eps) := \sup_{q\in Q} \sup_{|u-u_{0,q}| \leq \eps}|g_q'(u_{0,q}) - g_q'(u)|$ and assume $\chi(\eps) \to 0$ as $\eps \to 0$. Consider a collection of functions $g_{n,q}:[0,1]\to\R, q\in Q$ that satisfies for any $\eps_n \to 0$
\beq\label{eq:cpsi1}
\xi_n(\eps_n) := \sup_{q\in Q}\sup_{|u-u_{0,q}|\leq\eps_n}|g_{n,q}(u_{0,q})-g_{n,q}(u) -(g_q(u_{0,q})-g_q(u))| = o(1),
\eeq   
that 
\beq\label{eq:cpsi2}
\sup_{q\in Q}\sup_{u \in [0,1]}|g_{n,q}(u)-g_q(u)| = o(1),
\eeq
and define
\beq\label{eq:cpsi3}
R_n := \sup_{q\in Q} \sup_{|u-u_{0,q}|\leq\delta} |g_{n,q}(u)-g_q(u)|.
\eeq
Then there exists a constant $C_0$ depending on $g_{min}, \chi$ only such that for any collection of points $a_q,b_q \in (0,1)$ with $u_q-\delta < a_q+\eps \leq u_q \leq b_q - \eps < u_q + \delta$ with $\eps>0$ fixed we have for sufficiently large $n$
\beq \label{eq:psi1}
\Psi_{g_{n,q}}(0) = a_q + \int_{a_q}^{b_q} I\{g_{n,q}(u)\leq 0\} du \quad \forall q\in Q
\eeq 
and 
\beq \label{eq:psi2}
\sup_{q\in Q} \Big|\Psi_{g_q}(0) - \Psi_{g_{n,q}}(0) + \frac{g_{n,q}(u_{0,q})}{g_q'(u_{0,q})}\Big| \leq \frac{2\xi_n(C_0 R_n) + 4C_0 R_n \chi(C_0 R_n)}{\inf_{q\in Q} g_q'(u_{0,q})}.
\eeq
\end{theo}

\bigskip

Properties of the rearrangement map in a statistical context have been previously considered by \cite{DeNePi2006,neumeyer2007,DeVo2008,ChFeGa2010,vobidene2013}, among others. In particular, \cite{DeNePi2006,neumeyer2007,DeVo2008,vobidene2013} considered smoothed versions of the rearrangement map, and their results are not applicable in our setting. To the best of our knowledge, the only work that provides general results which can be used to obtain a Bahadur representation for the unsmoothed rearrangement map is the work by \cite{ChFeGa2010}, hereafter CFG. The result in CFG which is closest to Theorem \ref{theo:rear} is Proposition 2. In Proposition 2, CFG consider properties of the map $\Psi$ for functions $g$ which are allowed to take the value zero in several points on $[0,1]$ and are continuously differentiable. In this respect, Theorem \ref{theo:rear} is less general as we only allow for functions $g$ that cross zero at a unique point $u_0 \in (0,1)$. However, for non-decreasing functions with only one zero crossing, the assumptions of Theorem \ref{theo:rear} are weaker than those of Proposition 2 in CFG. More precisely, in Remark~\ref{rem:rearr} we shall show that the results in the first part of Proposition~2 of CFG can be recovered from our Theorem~\ref{theo:rear}. Additionally, below we provide a simple example where Theorem~\ref{theo:rear} is applicable but the assumptions of Proposition~2 in CFG fail -- see Example~\ref{ex:cfg}.

\begin{rem}{\rm \label{rem:rearr}
In this Remark we shall demonstrate that the first part of Proposition 2 in CFG can be obtained from Theorem \ref{theo:rear} under certain assumptions. Recall the setting of Proposition 2 in CFG -- we have a function $Q: (0,1)\times \Xc \to \R$ that is continuously differentiable in both arguments [see CFG, Assumption 1]. For uniformly bounded functions $h_n: (0,1)\times \Xc \to \R$ CFG define the map $F(y|x,h_n) := \int_0^1 I\{Q(u|x) + t_n h_n(u|x) \leq y\} du$. Note that $F(y|x,h_n) = \Psi_{g_{(x,y),n}}(0)$ where $g_{(x,y),n}(u) := Q(u|x) + t_n h_n(u|x) - y$. Similarly, the function $F(y|x) := \int_0^1 I\{Q(u|x) \leq y\}du$ defined in equation (2.4) of CFG can be represented as $F(y|x) = \Psi_{g_{(x,y)}}(0)$ where $g_{(x,y)}(u) := Q(u|x) - y$. In addition to the conditions of Proposition 2 of CFG we shall assume that the function $u\mapsto Q(u|x)$ is non-decreasing for every $x \in \Xc$. Under this assumption $Q(F(y|x)|x) = y$ for all $y$ with $\partial Q(u|x)/\partial u|_{u = F(y|x)} > 0$. In order to keep the presentation transparent, we consider a compact set of the form $S := [y_1,y_2]\times [x_1,x_2] \subset \Yc\Xc^*$ where $\Yc\Xc^*$ is as defined in CFG [the result can be generalized to arbitrary compact $S \subset \Yc\Xc^*$ at the cost of a more complex notation]. By the definition of $\Yc\Xc^*$ we have $u_{0,(x,y)} = F(y|x)$ and  
$g'_{(x,y)}(u_{0,(x,y)}) > 0$ for all $(x,y) \in S$. By continuity of $g'$ and compactness of $S$
it follows that $\inf_{(x,y)\in S} g'_{(x,y)}(u_{0,(x,y)}) > 0$.
We will now derive the conclusion in (2.5) of CFG from Theorem~\ref{theo:rear} applied with $Q = S$. Observe that for any $x,y$ we have $g_{(x,y)}(u) - g_{(x,y),n}(u) = t_n h_n(u|x)$. Thus \eqref{eq:cpsi2} holds and \eqref{eq:cpsi3} holds with $R_n = O(t_n)$. Moreover, $u_{0,(x,y)} = F(y|x)$ and for $\eps_n \to 0$
\bean
\xi_n(\eps_n) &=& t_n \sup_{(x,y) \in S} \sup_{|u|\leq \eps_n} \Big|h_n\Big(F(y|x) + u \Big|x\Big) - h_n\Big(F(y|x)\Big|x\Big)\Big|
\\
&=& o(t_n) + t_n \sup_{(x,y) \in S} \sup_{|u|\leq \eps_n}\Big|h\Big(F(y|x) + u \Big|x\Big) - h\Big(F(y|x)\Big|x\Big)\Big| = o(t_n)
\eean
since $h$ is continuous on $S$ and thus uniformly continuous since $S$ is compact. Hence all assumptions of Theorem~\ref{theo:rear} are satisfied and from \eqref{eq:psi2} we obtain
\[
\sup_{(x,y) \in S} \Big|\Psi_{g_{(x,y)}}(0) - \Psi_{g_{(x,y),n}}(0) + \frac{t_n h_n(F(y|x)|x)}{Q(F(y|x)|x)}\Big| = o(t_n).
\]
Since $h_n$ converges to $h$ uniformly this implies the statement (2.5) of CFG.  
\hfill $\blacksquare$ }
\end{rem}

\begin{example}\label{ex:cfg}{\rm
For simplicity we consider a set $Q$ with one element and drop the index $q$. Let $g(x) = x-1/2, g_n(x) := x - 1/2 + n^{-1/2}\sin(n^{1/4} x)$. Clearly $g$ satisfies the assumptions of Theorem \ref{theo:rear} and for the quantities defined in that theorem we have $u_0 = 1/2$, $\chi(\eps) \equiv 0$, $\xi_n(\eps_n) = \sup_{|v|\leq \eps_n} n^{-1/2}|\sin(n^{1/4}/2) - \sin(n^{1/4}/2 + n^{1/4}v)| = O(n^{-1/4}\eps_n)$, $R_n = O(n^{-1/2})$. Moreover, \eqref{eq:cpsi2} holds and thus \eqref{eq:psi2} yields the representation
\[
\Psi_{g_q}(0) - \Psi_{g_{n,q}}(0) = - n^{-1/2}\sin(n^{1/4}/2) + O(n^{-3/4}).
\]
In order to apply Proposition 2 of CFG we would need to show that $\sin(n^{1/4} x)$ converges uniformly to a continuous function, and this is obviously not the case. \hfill $\blacksquare$
}
\end{example}

Summarizing, Proposition 2 in CFG and Theorem \ref{theo:rear} in the present paper are valid under different sets of assumptions and none of the results implies the other one. Proposition 2 in CFG is particularly useful in settings where a functional central limit theorem holds. Theorem \ref{theo:rear} requires stronger assumptions on $g$ but can be applied without such an assumption.  

\section{Comments on practical implementation and a brief simulation study} \label{sec:sim}

In this section, we briefly comment on some practical aspects of implementing the proposed estimators $\hat p_w(a,b)$ and demonstrate their properties in two numerical examples. Several choices need to be made when constructing $\hat p_w(a,b)$ in practice: the values for $a,b$, the predictor $Z$ with corresponding coefficient normalized to one, and the bandwidth $h_n$.  

For selecting $a,b$, we suggest to start with $a = .01, b= .99$. In most cases of practical relevance, this will ensure that $p_w$ is estimated consistently as long as $.01 < p_w < .99$. If more `extreme' choice probabilities are of interest, $a,b$ can be adjusted. Note, however, that for realistic sample sizes it can be difficult to estimate such an `extreme' choice probability accurately. In simulations, the integral in~\eqref{hatpw} can be approximated by a Riemann sum over a uniformly spaced grid of quantiles on $[a,b]$. Some simulation evidence on the impact of choosing $a,b$ is given in the last paragraph of this section. 

For the choice of coefficient to normalize, there are two possible scenarios. If prior information about a coefficient that is likely to be non-zero and have the same sign for all  relevant values of $\tau$ is available, this coefficient should be chosen. If there is no such information, we propose to run several regressions with the same data, normalizing a different coefficient to have absolute value $1$ each time. If for one of the settings the sign is estimated to the same across all values of $\tau$, this normalization should be selected. If this is not the case for any of the normalizations, this indicates that assumption (A) might be violated since each of the coefficients changes signs at some point. This seems highly unlikely in practice since it means that the sign of the effect of \textit{each} predictor depends on the quantile. If the sign of a coefficient is different for only a small proportion of quantiles in the quantile grid, this might be due to estimation error. In practice, we suggest to normalize the coefficient that has the highest proportion of signs estimated to be the same across the quantile grid that is used to approximate the integral in~\eqref{hatpw}.

In order to select the bandwidth $h_n$, we propose to use $K$-fold cross-validation~(see \cite{friedman2001}, Chapter 7.10). The cross-validation procedure is described below. 

\begin{enumerate}
\item Fix a grid of candidate bandwidth parameters $h_1<...<h_N$ and a grid of quantile values $\Tc \subset [a,b]$.
\item Randomly divide $\{1,...,n\}$ into $K$ disjoint, (approximately) equally sized subsets $B_1,...,B_K \subset \{1,...,n\}$.
\item For $k=1,...,K$, $j = 1,...,N$ and $\tau \in \Tc$ compute the estimators $\hat\beta_{k,j}(\tau)$ from the data $(W_i,Y_i)_{i \in B_k^C}$ with $B_k^C := \{1,...,n\} \backslash B_k$ using the bandwidth $h_j$.
\item For each $i \in B_k$ compute $\hat p_{W_i}$ from $\{\hat\beta_{k,j}(\tau)\}_{\tau \in \Tc}$ and define for a pre-specified set $\Wc_0 \subset \Wc$
\begin{equation}\label{eq:cv}
\hat CV(j) := \sum_{k=1}^K \sum_{i \in B_k} (\hat p_{W_i} - Y_i)^2I\{W_i \in \Wc_0\} 
\end{equation} 
\item Let $j_{min} := \arg\min_{j=1,...,N} \hat CV(j)$ and set the bandwidth to $h_{j_{min}}$.
\end{enumerate} 

\begin{rem}
{\rm
For a motivation of the cross-validation criterion in~\eqref{eq:cv}, note that by independence between $\hat p_{W_i}$ and $Y_i$ for $i \in B_k$, we have
\begin{eqnarray*}
&&\E[(\hat p_{W_i}-Y_i)^2I\{W_i \in \mathcal{W}_0\}] 
\\
&=& \E[\E[(\hat p_{W_i}-Y_i)^2I\{W_i \in \mathcal{W}_0\}|W_i]]
\\
&=& \E\Big[Var(\hat p_{W_i}-Y_i|W_i)I\{W_i \in \mathcal{W}_0\} + \Big(\E[\hat p_{W_i}-Y_i|W_i]\Big)^2I\{W_i \in \mathcal{W}_0\} \Big]
\\
&=& \E\Big[Var(\hat p_{W_i}|W_i)I\{W_i \in \mathcal{W}_0\} + \Big(\E[\hat p_{W_i}|W_i]-p_{W_i}\Big)^2I\{W_i \in \mathcal{W}_0\} + Var(Y_i|W_i)I\{W_i \in \mathcal{W}_0\}\Big]
\\
&=& \int_{\mathcal{W}_0} \E[(\hat p_w - p_w)^2] dP_W(w) + \E[Var(Y_i|W_i)I\{W_i \in \mathcal{W}_0\}]
\end{eqnarray*}
The first two terms above correspond to the (weighted according to the distribution of $W$, and integrated over $\mathcal{W}_0$) mean squared error of the estimator $\hat p_{W_i}$ while the last term does not depend on $\hat p_{W_i}$ and thus the bandwidth. Thus, minimizing the expression above with respect to the bandwidth $h_n$ which enters $\hat p_{W_i}$, corresponds to finding a bandwidth which minimizes a weighted version of the mean squared error of the estimator $\hat p_{W_i}$. In practice $\E[(\hat p_{W_i}-Y_i)^2I\{W_i \in \mathcal{W}_0\}] $ is not available, and cross-validation aims to estimate this term by a plug-in version which is given in~\eqref{eq:cv}.   
} \hfill $\blacksquare$
\end{rem}

\begin{rem}
{\rm
As pointed out by a referee cross-validation usually aims at producing a bandwidth which balances bias and variance. If bias is a major concern under-smoothing (i.e. selecting a smaller bandwidth) can be used, although this will increase the variance. An approach that is simple to implement would be to obtain the bandwidth determined by cross-validation in a preliminary step and make that bandwidth smaller when computing the final estimator.   
} \hfill $\blacksquare$
\end{rem}

Next, we illustrate the properties of the estimated probabilities $\hat p_w$ in a small simulation study. To the best of our knowledge, the only article in the literature that considers estimation of conditional choice probabilities without assuming a specific single index or even parametric model is~\cite{khan2013}. While that reference provides no theory on properties of resulting choice probability estimators, some simulation evidence for those estimators is reported there and we are going to compare the properties of the proposed estimators with those of~\cite{khan2013}. To this end, we consider the following data-generation process from~\cite{khan2013} [see also \cite{horowitz1992}]
\begin{align*}
Y_i = I\{W_{1,i} + W_{2,i} + V_i \geq 0\}
\end{align*}   
where $W_{1,i}\sim \Nc(0,1), W_{2,i}\sim \Nc(1,1)$ and the $V_i$ follow four different distributions
\begin{enumerate}
\item Design 1: $V_i$ independent of $W$, logistic with median $0$ and variance $1$.
\item Design 2: $V_i$ independent of $W$, uniform with median $0$ and variance $1$.
\item Design 3: $V_i$ independent of $W$, $t$ with three degrees of freedom scaled to have variance $1$.
\item Design 4: $V_i = U_i (1 + (W_{1,i} + W_{2,i})^2)^2 $, $U_i$ independent of $W$, logistic with median $0$ and variance $1$.
\end{enumerate}
Note that Design 1-3 satisfies assumption (M) while this assumption is violated in Design 4. Following~\cite{khan2013}, we normalize the coefficient in front of $W_{1,i}$ to have absolute value one. The estimator $\hat p_w$ is computed based on the following algorithm

\begin{enumerate}
\item \textbf{Input}: parameters $0<a<b<1$, grid size $G$, point of interest $w$.
\item Define grid points $\tau_g := a + (b-a)/(2G) + (g-1)*(b-a)/G$, $g=1,...,G$ and for each $g=1,...,G$ compute $\hat \beta_{\tau_g} = (\hat s_{\tau_g}, \hat b_{\tau_g})$ as the solution of~\eqref{eq:esti}. Here, first optimize over $b$ for $s=1, s=-1$ separately and later take the solution which leads to the bigger value of $\hat S_{n,\tau}(s,b)$. \footnote{ In the simulations in the present paper we use the general purpose optimization routine \texttt{optim} in \texttt{R} Core Team (2015), version 3.2.2 at default settings and starting value $b = (0,1)$ to optimize~\eqref{eq:esti} over $b$.}
\item Compute $\hat p_w(a,b)$ by approximating the integral in~\eqref{hatpw} through a sum\footnote{in the present simulation study, we set $G=50$}, more precisely
\[
\hat p_w(a,b) := 1-b + \frac{b-a}{G}\sum_{g=1}^G I\{w^T \hat \beta_{\tau_g} \geq 0\}.
\]
\item \textbf{Output}: $\hat p_w(a,b)$.
\end{enumerate}    

All of the following results are based on $5-$fold cross validation with $500$ simulation repetitions. The function $\Kc$ was chosen to be the Kernel $K_4$ given on page 516 in~\cite{horowitz1992} but scaled so that the support of its derivative is $[-1,1]$. The grid of candidate bandwidth values was set to $h_j = 4jn^{-1/7}, j=1,...,20$ (note that the MSE-optimal bandwidth should be proportional to $n^{-1/7}$ since $K_4$ is a Kernel of order 4) where $n$ denotes the total sample size. The set $\Wc_0$ in the cross validation was chosen to be $[-3,3]^2$. 

Table~\ref{tab1} summarizes the average asymptotic AMSE for $\hat p_w$, where the MSE is averaged over a uniformly spaced $50$ by $50$ grid on $[-3,3]$ for the predictors $W_1,W_2$ [this is exactly the approach taken by~\cite{khan2013} and the AMSE values for the sieve estimator are taken from~\cite{khan2013}]. In Design 1-3 both procedures perform reasonably well. The approach of~\cite{khan2013} performs better for a sample size of $n = 250$, while for $n = 500, 1000$ the procedure proposed here has a slight edge.\footnote{The fact that the AMSE for the sieve estimators from~\cite{khan2013} tends to increase from $n = 250$ to $n = 500$ is probably due to the choice of sieve basis which has more elements for $n = 500, 1000$ compared to $n = 250$.} In Design 4, both procedures perform worse than in Design 1-3. Given the rather complex shape of the true choice probabilities (cf Figure~\ref{fig4}) this is not too surprising.

Since Design 1-3 are similar, the remaining discussion will focus on Design 1 and 4 while details on Design 2 and 3 are given in the Appendix (see Section~\ref{sec:addsim}). The true choice probabilities, along with the estimated choice probabilities (averaged over $500$ simulation repetitions) are presented in Figure~\ref{fig1} and Figure~\ref{fig4}, respectively. Figure~\ref{fig1} shows that the procedure proposed in the present paper is, on average, able to reproduce the overall shape of the conditional choice probability (as a function of $w$) fairly well and this is true even for $n = 250$. Similar comments apply to the approach of~\cite{khan2013}, c.f. Figure 1 in the latter paper. 

Figure~\ref{fig4} in the present paper and Figure 4 in~\cite{khan2013} indicate that both procedures suffer from a substantial bias. Overall, the shape is reproduced better by the sieve procedure in~\cite{khan2013} which is expected since asymptotically the sieve should be able to consistently estimate the choice probabilities while this is not the case for our approach. Interestingly, the average MSE of our procedure is lower than that of~\cite{khan2013} for the sample sizes considered in this simulation. This indicates that sometimes completely non-parametric procedures can only develop their full advantage when the sample size is substantial. 

In the final part of this simulation study, we analyse the effect of $a,b$ on estimated choice probabilities. To this end, we consider the three covariate values $(-0.5, -0.5), (0,0)$ and $(0.5,0.5)$ along with all combinations of $1-a \in \{0.01, 0.15, 0.25\}, 1-b \in \{0.75, 0.85, 0.99\}$. Results for Design 1 and 4 are reported in Table~\ref{tab2} and Table~\ref{tab5} while the corresponding results for Design 2 and 3 are deferred to Section~\ref{sec:addsim} in the Appendix. A close look at both tables reveals that for all choices of $1-a,1-b$ considered there is no big impact on estimation of $p_{w}$ for $w = (0,0)^T$. Additionally, only the choice of $a$ has an impact on estimating $p_{w}$ for $w = (-0.5,-0.5)^T$ (in this case $p_w$ is smaller than $0.5$) and only the choice of $b$ impacts the estimation of $p_{w}$ for $w = (0.5,0.5)^T$ (in this case $p_w$ is greater than $0.5$). This is in line with the theory.  

\begin{table} 
\begin{center}
\begin{tabular}{c||c|c|c|}
& n = 250 & n = 500 & n = 1000 \\
\hline
Design 1 & 0.0097 (0.0062) & 0.0057 (0.0065) & 0.0032 (0.0031)\\
\hline
Design 2 & 0.0103 (0.0070) & 0.0061 (0.0077) &  0.0036 (0.0041)\\
\hline
Design 3 & 0.0082 (0.0081) &  0.0050 (0.0085) & 0.0028 (0.0057)\\
\hline
Design 4 & 0.0511 (0.0692) & 0.0435 (0.0617) & 0.0391 (0.0600)\\
\hline
\end{tabular}
\end{center}
\caption{Average MSE for estimating choice probabilities. Average is taken over an equispaced $50$ by $50$ grid over $[-3,3]^2$. Values in brackets correspond to the average MSE of the sieve estimators considered in~\cite{khan2013} and are taken from that reference.}	 \label{tab1}
\end{table}


\begin{table} \small
\begin{tabular}{cc||c|c|c|c|c|c|c|c|c}
  & & 1-a~~1-b &1-a~~1-b&1-a~~1-b&1-a~~1-b&1-a~~1-b&1-a~~1-b&1-a~~1-b&1-a~~1-b&1-a~~1-b\\
  & & 0.01~0.99&0.15~0.99&0.25~0.99&0.01~0.85&0.15~0.85&0.25~0.85&0.01~0.75&0.15~0.75&0.25~0.75\\
	\hline
	$p_w$&  &\multicolumn{9}{c}{n = 250 }
	\\
	\hline
0.14 &RMSE& 9.03& 7.83& 12.13& 8.99& 7.8& 12.12& 8.95& 7.77& 12.12\\
&bias& 2.12& 5.3& 11.85& 2.07& 5.28& 11.85& 2.02& 5.26& 11.84\\
\hline
0.5 & RMSE& 10.06& 10.03& 9.89& 9.99& 9.95& 9.79& 9.73& 9.67& 9.48\\
&bias& 0.33& 0.49& 0.68& 0.15& 0.31& 0.52& -0.08& 0.07& 0.28\\
\hline
0.86& RMSE& 6.47& 6.43& 6.39& 6.16& 6.13& 6.1& 11.35& 11.34& 11.34\\
&bias& -2.91& -2.85& -2.8& -4.49& -4.47& -4.46& -11.28& -11.28& -11.27\\
\hline
&\multicolumn{9}{c}{n = 500 }
	\\
\hline
0.14 &RMSE& 7.11& 6.08& 11.39& 7.07& 6.06& 11.39& 7.04& 6.03& 11.39\\
&bias& 1.69& 4.24& 11.32& 1.64& 4.23& 11.32& 1.6& 4.21& 11.32\\
\hline
0.5& RMSE& 7.6& 7.58& 7.52& 7.57& 7.54& 7.47& 7.49& 7.45& 7.36\\
&bias& 0.21& 0.37& 0.53& 0.04& 0.21& 0.38& -0.12& 0.03& 0.2\\
\hline
0.86& RMSE& 4.82& 4.78& 4.75& 4.72& 4.7& 4.69& 11.02& 11.02& 11.02\\
&bias& -2.46& -2.41& -2.36& -3.63& -3.62& -3.61& -11.01& -11.01& -11.01\\
\hline
&\multicolumn{9}{c}{n = 1000 }
	\\
\hline
0.14 & RMSE&  5.37& 5.07& 11.07& 5.34& 5.05& 11.07& 5.3& 5.03& 11.07\\
& bias& 2.23& 3.78& 11.06& 2.18& 3.76& 11.06& 2.14& 3.75& 11.06\\
\hline
0.5& RMSE& 5.09& 5.09& 5.09& 5.07& 5.05& 5.04& 5.05& 5.02& 5\\
&bias& 0.38& 0.55& 0.7& 0.22& 0.38& 0.54& 0.07& 0.22& 0.38\\
\hline
0.86& RMSE& 3.63& 3.59& 3.56& 3.65& 3.64& 3.63& 10.99& 10.99& 10.99\\
&bias& -1.93& -1.88& -1.83& -2.88& -2.87& -2.86& -10.99& -10.99& -10.99\\
\hline
\end{tabular} 
\caption{\small Estimation quality of conditional choice probabilities for different values of $a,b$. Root mean squared error (RMSE) and bias are multiplied by $100$. Data are generated from Design 1.}	 \label{tab2}
\end{table}

\begin{table} \small
\begin{tabular}{cc||c|c|c|c|c|c|c|c|c}
  & & 1-a~~1-b &1-a~~1-b&1-a~~1-b&1-a~~1-b&1-a~~1-b&1-a~~1-b&1-a~~1-b&1-a~~1-b&1-a~~1-b\\
  & & 0.01~0.99&0.15~0.99&0.25~0.99&0.01~0.85&0.15~0.85&0.25~0.85&0.01~0.75&0.15~0.75&0.25~0.75\\
	\hline
	$p_w$&  &\multicolumn{9}{c}{n = 250 }
	\\
	\hline
0.14& RMSE &6.99& 6.72& 11.48& 6.9& 6.64& 11.44& 6.85& 6.59& 11.43\\
&bias& 3.32& 4.96& 11.38& 3.24& 4.91& 11.35& 3.18& 4.89& 11.34\\
\hline
0.5& RMSE &11.87& 11.8& 11.56& 11.75& 11.67& 11.4& 11.57& 11.46& 11.17\\
&bias& -0.06& 0.11& 0.38& -0.25& -0.08& 0.2& -0.47& -0.31& -0.03\\
\hline
0.86& RMSE &11.92& 11.84& 11.77& 12.13& 12.08& 12.03& 13.35& 13.33& 13.32\\
&bias& -10.89& -10.81& -10.73& -11.22& -11.17& -11.12& -13.06& -13.05& -13.03\\
\hline
&  &\multicolumn{9}{c}{n = 500 }
	\\
\hline
0.14& RMSE &5.46& 5.55& 11.08& 5.41& 5.53& 11.08& 5.37& 5.51& 11.07\\
&bias& 3.61& 4.39& 11.06& 3.55& 4.37& 11.06& 3.5& 4.35& 11.06\\
\hline
0.5& RMSE &9.49& 9.46& 9.27& 9.46& 9.41& 9.2& 9.42& 9.36& 9.13\\
&bias& 0.05& 0.21& 0.42& -0.12& 0.05& 0.26& -0.27& -0.12& 0.1\\
\hline
0.86& RMSE &11.92& 11.84& 11.77& 12.15& 12.1& 12.05& 13.08& 13.07& 13.06\\
&bias& -11.31& -11.22& -11.15& -11.57& -11.52& -11.47& -12.89& -12.87& -12.86\\
\hline
&  &\multicolumn{9}{c}{n = 1000 }
	\\
\hline
0.14& RMSE &5.02& 5.22& 10.99& 4.97& 5.2& 10.99& 4.92& 5.18& 10.99\\
&bias& 4.03& 4.45& 10.99& 3.97& 4.44& 10.99& 3.92& 4.42& 10.99\\
\hline
0.5& RMSE &6.36& 6.35& 6.33& 6.34& 6.31& 6.29& 6.33& 6.28& 6.24\\
&bias& 0.05& 0.21& 0.36& -0.12& 0.05& 0.21& -0.27& -0.11& 0.05\\
\hline
0.86& RMSE &11.96& 11.88& 11.8& 12.19& 12.14& 12.09& 12.94& 12.93& 12.92\\
&bias& -11.57& -11.49& -11.41& -11.82& -11.77& -11.72& -12.82& -12.81& -12.8\\
\hline
\end{tabular} 
\caption{\small Estimation quality of conditional choice probabilities for different values of $a,b$. Root mean squared error (RMSE) and bias are multiplied by $100$. Data are generated from Design 4.}	\label{tab5}
\end{table}

\begin{center}

\begin{figure}
\includegraphics[width = 3.5in]{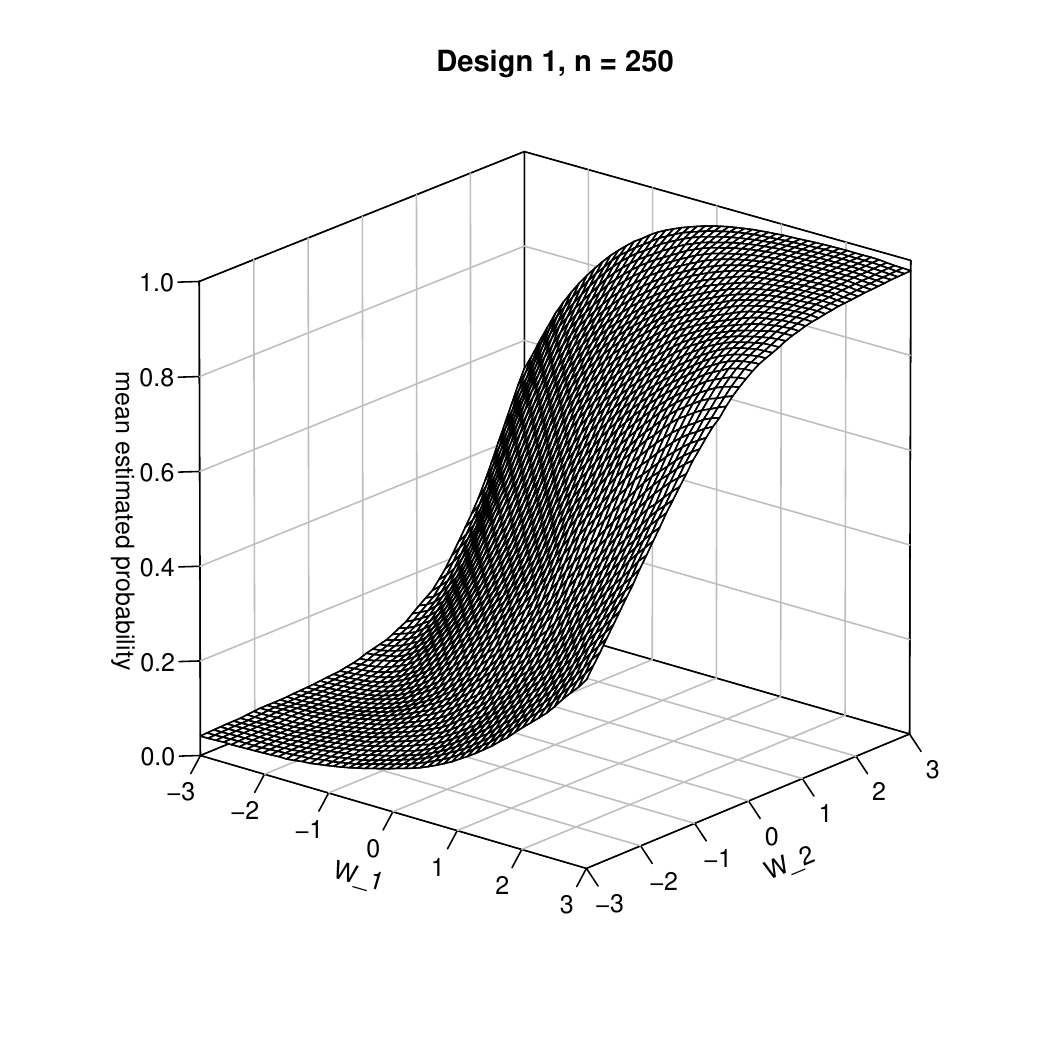}
\includegraphics[width = 3.5in]{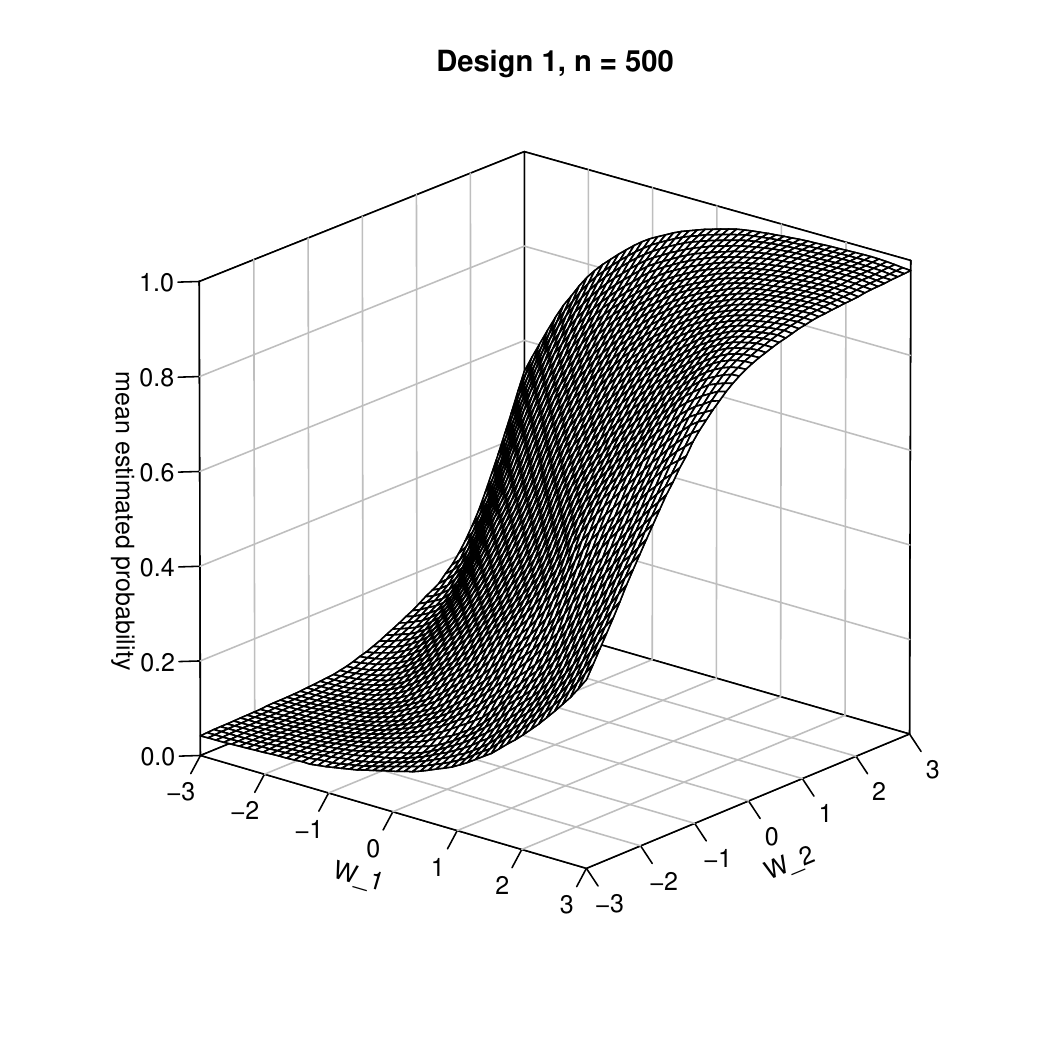}
\includegraphics[width = 3.5in]{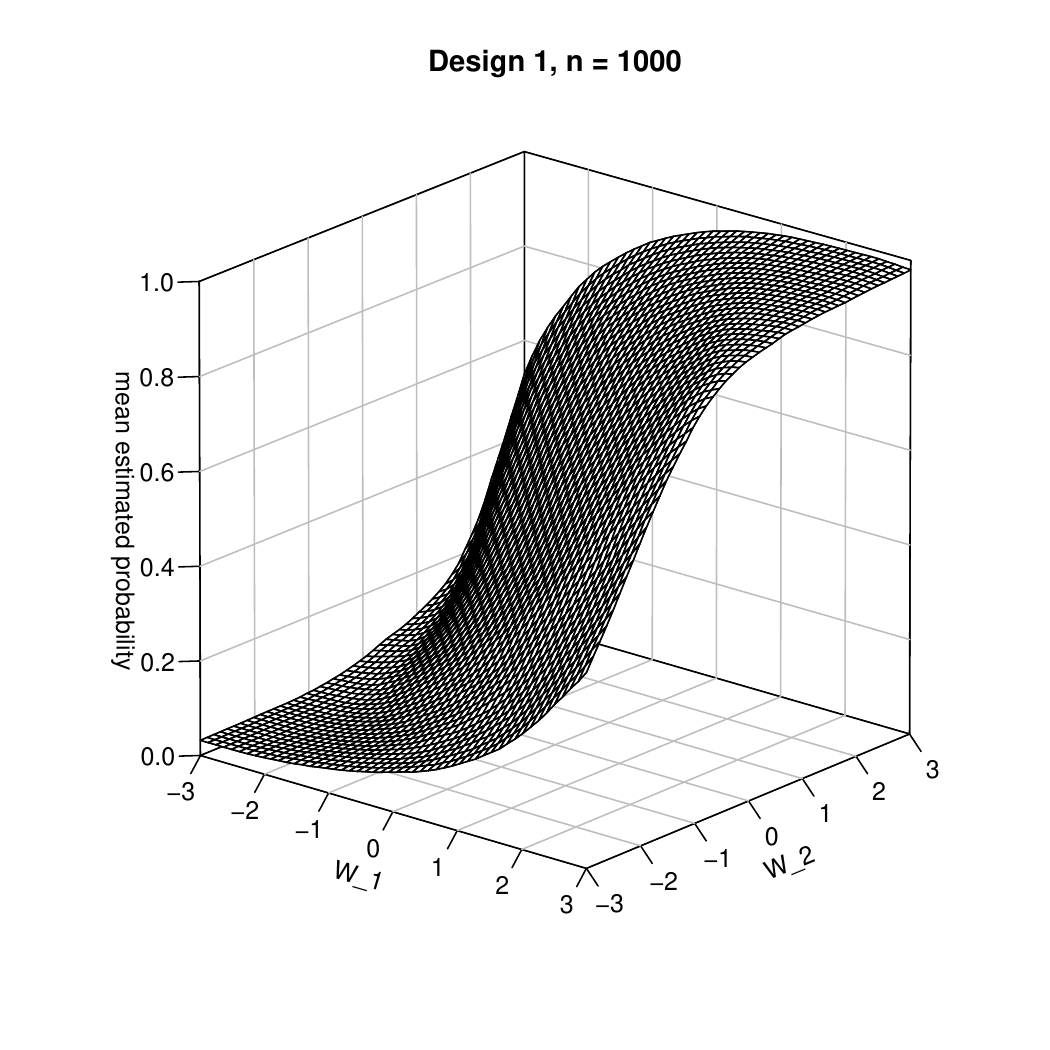}
\includegraphics[width = 3.5in]{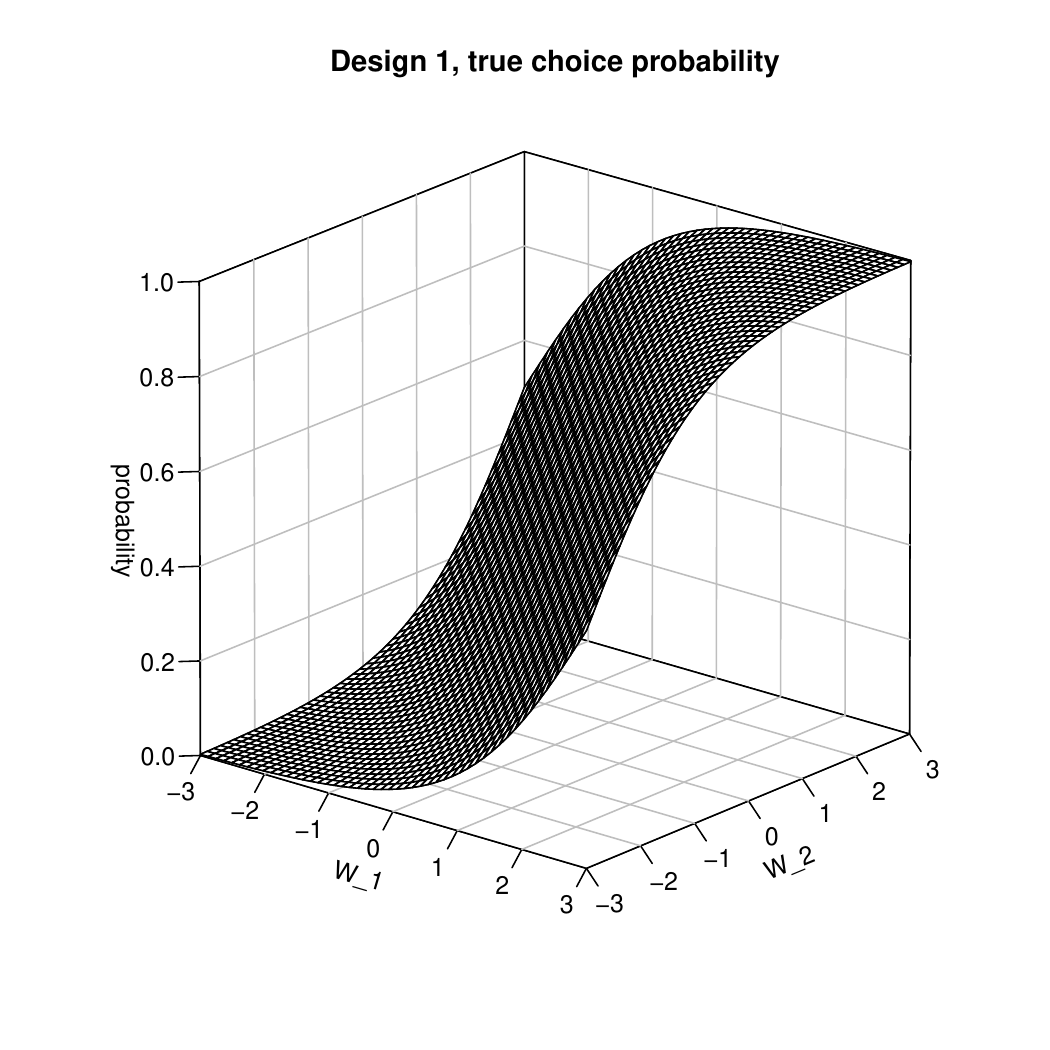} 
\caption{Average (over 500 simulations) estimated (first 3 figures) and true (bottom right) choice probabilities as functions of covariates in Design 1 } \label{fig1}
\end{figure}

\begin{figure}
\includegraphics[width = 3.5in]{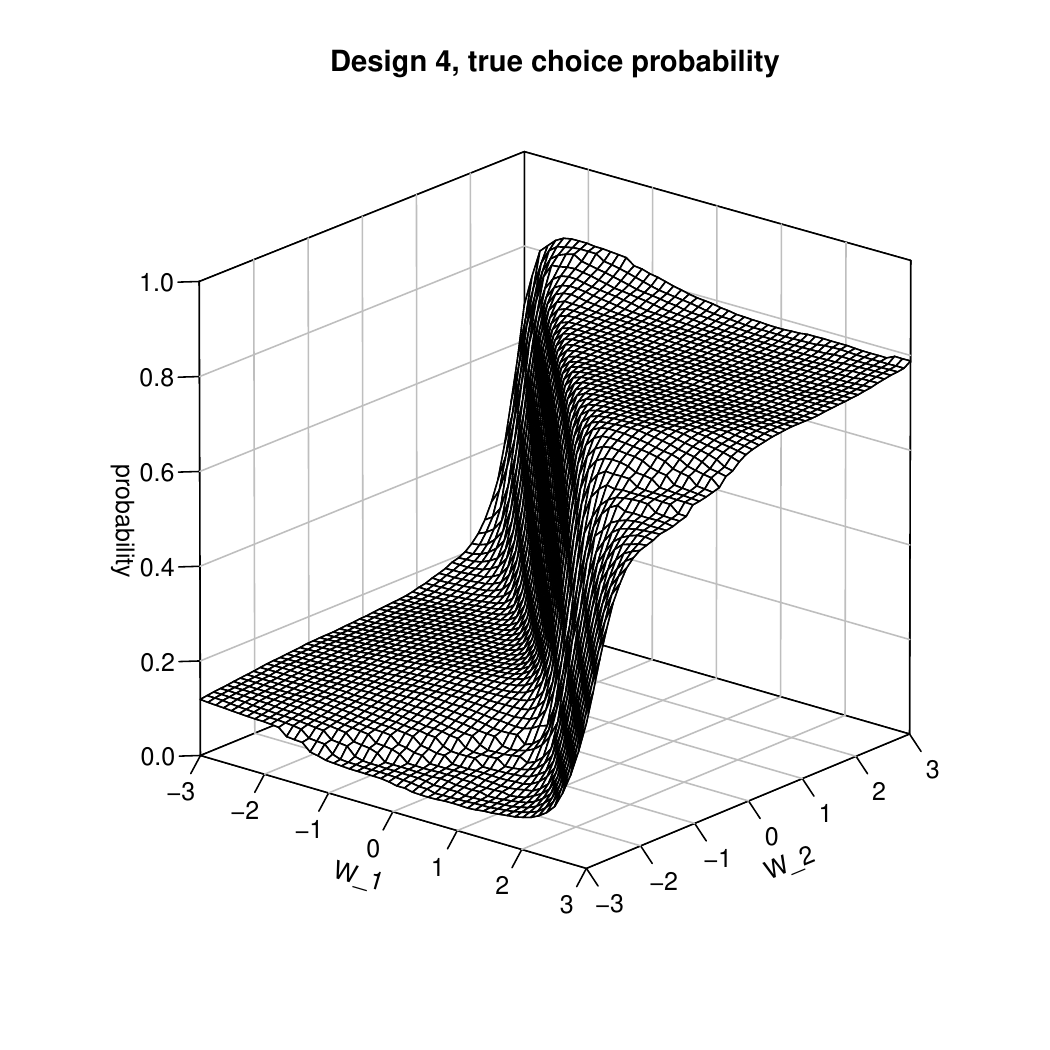}
\includegraphics[width = 3.5in]{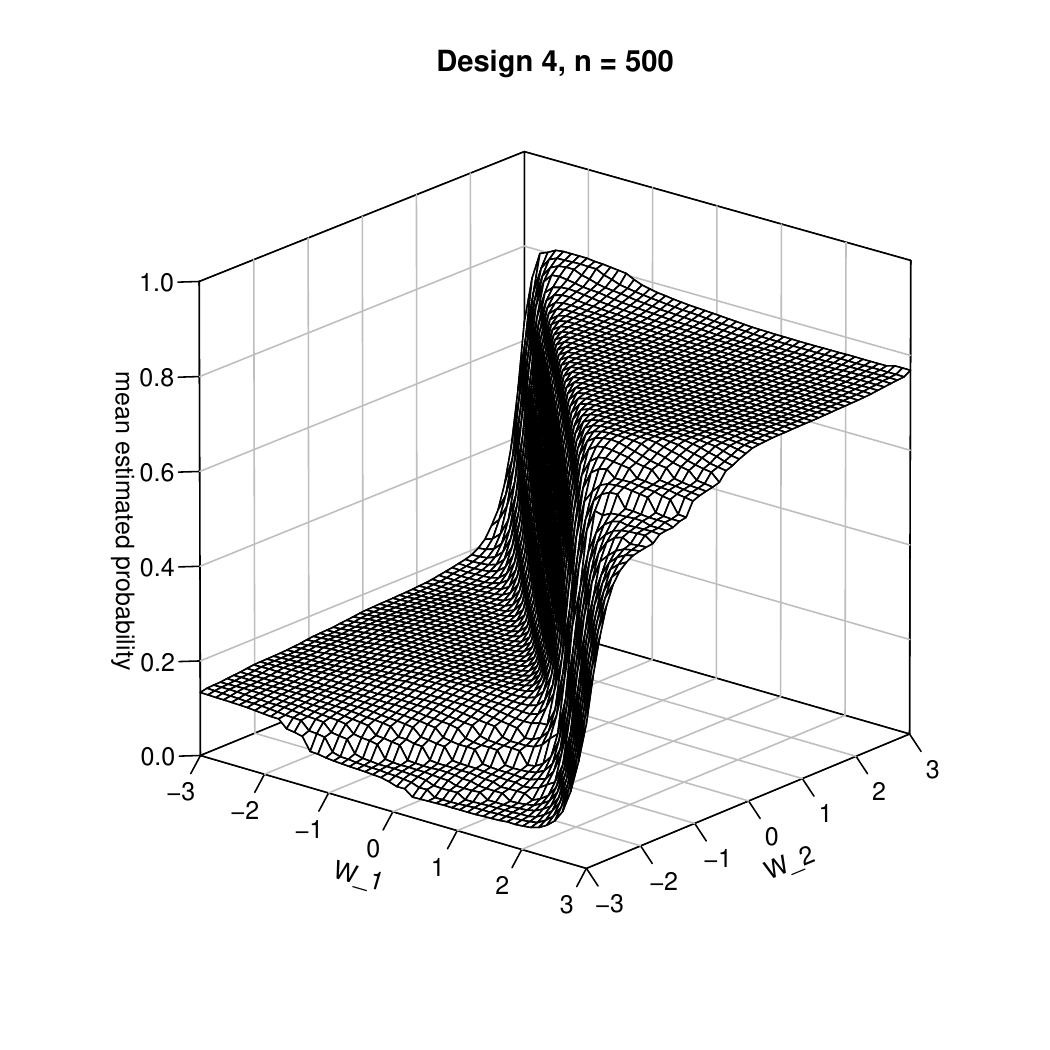}
\includegraphics[width = 3.5in]{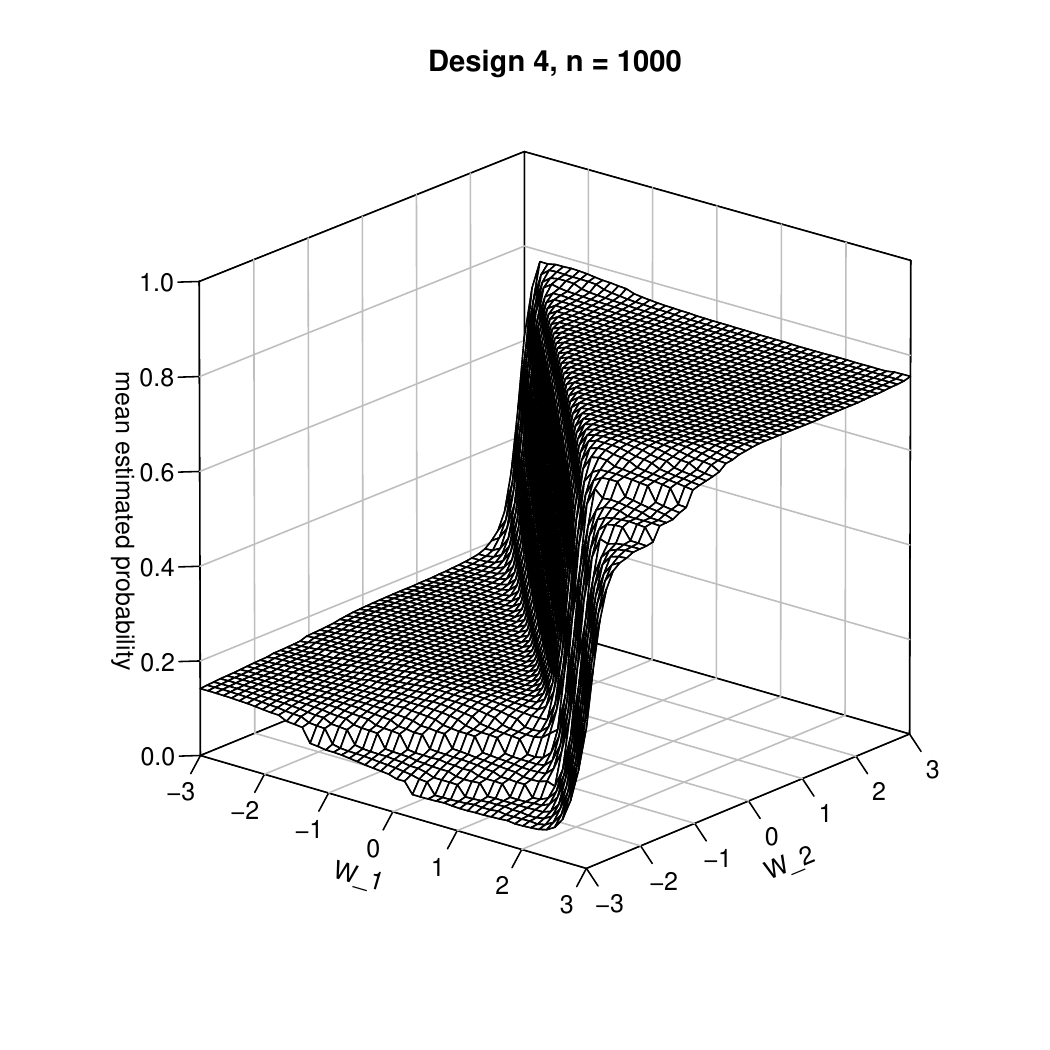}
\includegraphics[width = 3.5in]{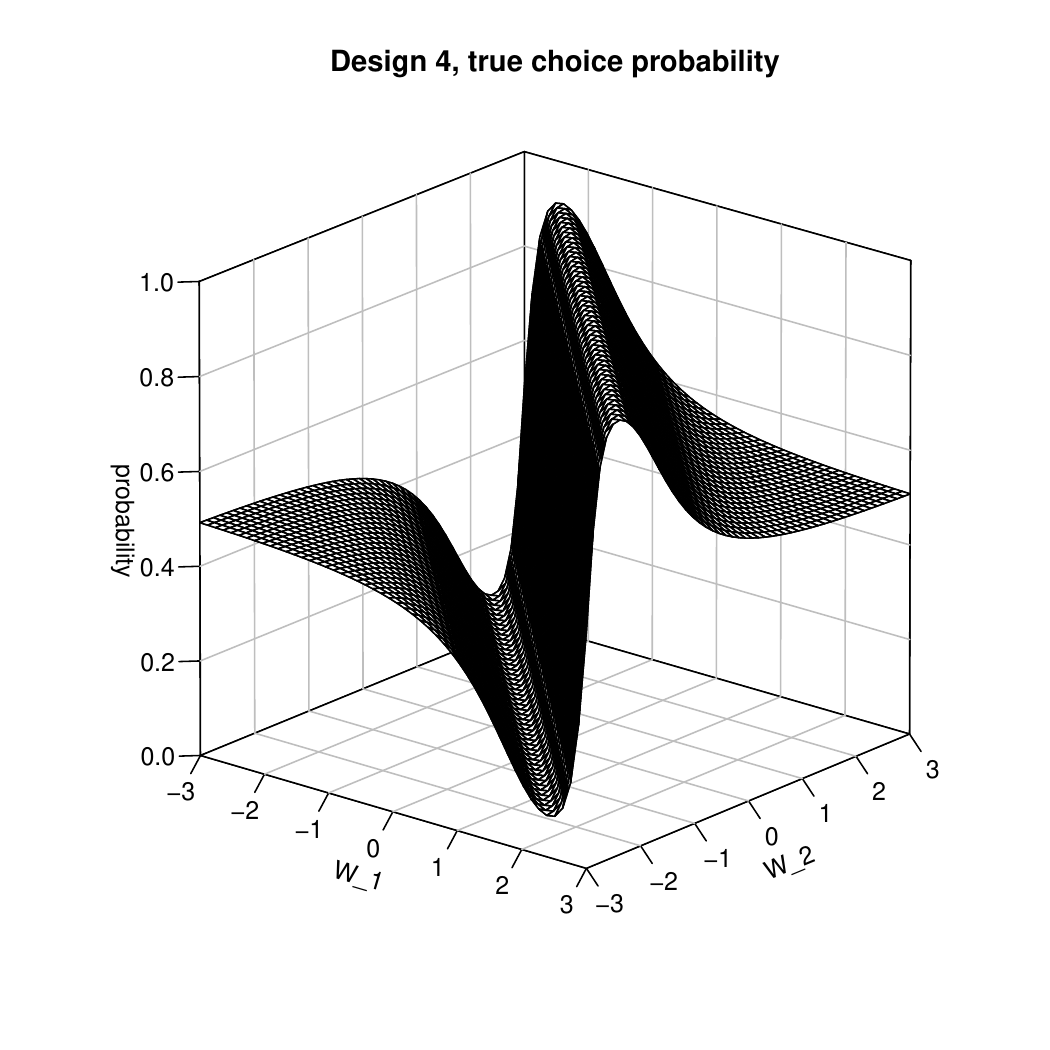} 
\caption{Average (over 500 simulations) estimated (first 3 figures) and true (bottom right) choice probabilities as functions of covariates in Design 4 } \label{fig4}
\end{figure}
\end{center}

\newpage

\section{Appendix: Proofs}

\subsection{Proofs for Section~\ref{sec:coef} }

We begin with some useful algebraic identities. First, we prove that under assumptions (A), (D1)-(D5), (M) and (Q)
\beq \label{eq:Fprime}
\frac{\partial}{\partial u}\Big(F_{Y^*|X,Z}(0|x,s_0(-x^T\bar b_\tau+u))\Big)\Big|_{u=0} = -|\beta_{\tau,1}| f_{Y^*|X,Z}(0|x,-x^Ts_0\bar b_\tau).
\eeq 
To this end, note that $y(u) := F_{Y^*|X,Z}(0|x,s_0(-x^T\bar b_\tau+u))$ satisfies the equation (for $u$ sufficiently small)
\[
q_{y(u)}((s_0(-x^T\bar b_\tau+u),x^T)^T) = s_0(-x^T\bar b_\tau+u) \beta_{y(u),1} + x^T b_{y(u)} = u s_0 \beta_{y(u),1} + x^T b_{y(u)} - s_0\beta_{y(u),1}x^T\bar b_\tau,
\]
this follows from the linearity of the conditional quantile function $q_\tau(w) = w^T\beta_\tau$. Additionally $y(0) = \tau$. Thus by the implicit function theorem $u \mapsto y(u)$ is differentiable and its derivative is given by
\[
- \frac{\frac{\partial}{\partial u}\Big(u s_0 \beta_{y,1} + x^T b_{y} - s_0\beta_{y,1}x^T\bar b_\tau\Big) \Big|_{y = \tau, u=0}}{\frac{\partial}{\partial y} \Big(u s_0 \beta_{y,1} + x^T b_{y} - s_0\beta_{y,1}x^T\bar b_\tau\Big) \Big|_{y = \tau, u=0}} = - s_0 \beta_{\tau,1}f_{Y^*|X,Z}(0|x,-s_0x^T\bar b_\tau).
\]
Here, the last step follows since
\[
\frac{\partial}{\partial y} \Big(u s_0 \beta_{y,1} + x^T b_{y} - s_0\beta_{y,1}x^T\bar b_\tau\Big) \Big|_{y = \tau, u=0} = \frac{\partial}{\partial y} \Big( -s_0x^T\bar b_\tau , x^T\Big)\beta_y \Big|_{y = \tau} = \frac{1}{f_{Y^*|X,Z}(0|x,-s_0x^T\bar b_\tau)}
\]
where we used the identities $(-s_0x^T\bar b_\tau , x^T)\beta_y = q_y((-s_0x^T\bar b_\tau , x^T)^T)$ and  $q_\tau((-s_0x^T\bar b_\tau , x^T)^T) = 0$, this shows \eqref{eq:Fprime}. Equation \eqref{eq:Fprime} implies that, under assumptions (D1)-(D5) and (Q),
\begin{multline} \label{eq:qprime}
\frac{\partial}{\partial u}\Big((\tau - F_{Y^*|X,Z}(0|x,s_0(-x^T\bar b_\tau+u)))f_{Z|X}(s_0(-x^T\bar b_\tau+u)|x) \Big)\Big|_{u=0} \\
= |\beta_{\tau,1}| f_{Y^*|X,Z}(0|x,-s_0x^T\bar b_\tau) f_{Z|X}(-s_0x^T\bar b_\tau|x).
\end{multline}


\textbf{Proof of (A), (M) and (S) implies (D1)} The proof is based on ideas from \cite{manski1985}. First, let us prove that under (A) and (S) we have for any fixed $\tau \in T$ 
\beq \label{eq:id1}
S_\tau((s,b^T)^T)-S_\tau(\beta_\tau) < 0 \quad \forall~ b\in B, s \in \{-1,1\}: (b,s) \neq (\bar b_\tau, s_0).
\eeq
To this end, recall the decomposition $S_\tau((s,b^T)^T)-S_\tau(\beta_\tau) = D_1((s,b^T)^T,\tau) + D_2((s,b^T)^T,\tau)$ where 
\bean
D_1((s,b^T)^T,\tau) &=& \E[(Y-(1-\tau))I\{sZ+X^Tb \geq 0> Zs_0 + X^T \bar b_\tau\}],
\\
D_2((s,b^T)^T,\tau) &=&  - \E[(Y-(1-\tau))I\{sZ+X^Tb <0 \leq Zs_0 + X^T \bar b_\tau\}]. 
\eean
Additionally, under (S) we have $E[Y-(1-\tau)|W=w] > 0$ if $w^T\beta_\tau > 0$ and $E[Y-(1-\tau)|W=w] < 0$ if $w^T\beta_\tau < 0$. This implies that $D_1((s,b^T)^T,\tau) \leq 0, D_2((s,b^T)^T,\tau) \leq 0$. Thus it suffices to prove that for $(s,b)\neq (s_0,\bar b_\tau)$ we have 
\beq \label{eq:id2}
P\Big(sZ+X^Tb > 0> Zs_0 + X^T \bar b_\tau\Big) + P\Big(sZ+X^Tb < 0 < Zs_0 + X^T \bar b_\tau\Big) > 0.
\eeq
To see this, begin by observing that under (S) we have $P(X^T b \neq  X^T \bar b_\tau) > 0$ if $b\neq \bar b_\tau$ [otherwise $P(X^T(b-\bar b_\tau)=0)=1$ which contradicts the assumption that $X$ is not concentrated on a proper subspace]. Assume without loss of generality that $P(X^T b <  X^T \bar b_\tau) > 0$. The assumptions on $Z$ imply that in this case also $P(sZ + X^T b < 0 < s_0 Z + X^T \bar b_\tau) > 0$, this can be proved by considering the cases $s=s_0, s= -s_0$ combined with the disjoint events $\{0 \leq X^T b < X^T \bar b_\tau\}$, $\{X^T b <  X^T \bar b_\tau \leq 0\}$, $\{X^T b < 0 < X^T \bar b_\tau \}$. For instance, if $s=s_0=1$ and $0 \leq x^T b <  x^T \bar b_\tau$ we have $sz + x^T b < 0 < sz + x^T \bar b_\tau$ for $z \in (-x^T \bar b_\tau,-x^T b)$, which by construction is an interval of positive length for every fixed $x$ satisfying $0 \leq x^T b <  x^T \bar b_\tau$. Thus $P(0 \leq X^T b < X^T \bar b_\tau)>0$ implies $P\Big(Z+X^Tb < 0 < Z + X^T \bar b_\tau\Big) > 0$. All other cases are handled similarly. Summarizing we have established that for $b\neq \bar b_\tau$ \eqref{eq:id2} holds. It remains to consider the case $b=\bar b_\tau, s \neq s_0$. From the conditions on $Z$, this case is obvious. This establishes \eqref{eq:id2}, and \eqref{eq:id1} follows.\\

Next, we shall prove (D1) by contradiction. Assume that (D1) does not hold. In that case there exists $\eps>0$ and sequences $(\tau_n)_{n\in \N}$ in $T$, $(b_n)_{n\in \N}$ in $B$, $(s_n)_{n\in \N}$ in $\{-1,1\}$ such that $S_{\tau_n}((s_n,b_n^T)^T)-S_{\tau_n}(\beta_{\tau_n}) \to 0$ and $\|(s_n,b_n) - (s_0,\bar b_{\tau_n})\| \geq \eps$. By compactness of $T,B$ there exists a subsequence $n_k$ and values $b^* \in B, \tau^* \in T, s^* \in \{-1,1\}$ such that $b_{n_k} \to b^*, s_{n_k}\to s^*, \tau_{n_k} \to \tau^*$ and $\|(s^*,b^*) - (s_0,\bar b_{\tau^*})\| \geq \eps$. By continuity of $(s,b,\tau) \mapsto D_1((s,b^T)^T,\tau) + D_2((s,b^T)^T,\tau)$ [this follows under (S) by majorized convergence and continuity of the distribution of $sZ+X^Tb$] we have
\[
0 = \lim_{n\to \infty} S_{\tau_n}((s_n,b_n^T)^T)-S_{\tau_n}(\beta_{\tau_n}) = S_{\tau^*}((s^*,(b^*)^T)^T)-S_{\tau^*}(\beta_{\tau^*}).
\]
However, since $\|(s^*,b^*) - (s_0,\bar b_{\tau^*})\| \geq \eps$, this contradicts \eqref{eq:id1}. Thus (D1) follows. \hfill $\Box$

\bigskip

\textbf{Proof of Lemma \ref{lem:unifcons}}
By Lemma 2.6.15 and Lemma 2.6.18 in \cite{vaarwell1996}, the classes of functions $\{(y,w) \mapsto yI\{w^T\beta \geq 0\}|\beta \in \R^{d+1}\}$ and $\{w \mapsto I\{w^T\beta \geq 0\}|\beta \in \R^{d+1}\}$ are VC-subgraph classes of functions. Together with Theorem 2.6.7 and Theorem 2.4.3 in the same reference this implies 
\[
\sup_{\tau\in[0,1],\beta\in \R^{d+1}} |S_\tau(\beta)-\tilde S_{n,\tau}(\beta)| = o(1) \quad a.s.
\]
To see this, note that by definition
\begin{multline*}
\sup_{\tau\in[0,1],\beta\in \R^{d+1}} |S_\tau(\beta)-\tilde S_{n,\tau}(\beta)| \leq 
\sup_{\beta\in \R^{d+1}} \Big|\frac{1}{n}\sum_i (Y_i I\{\beta^TW_i \geq 0\} - \E[Y_iI\{\beta^TW_i \geq 0\}]\Big|
\\
 + \sup_{\beta\in \R^{d+1}}\Big|\frac{1}{n}\sum_i (I\{\beta^TW_i \geq 0\} - P(\beta^TW_i \geq 0)\Big|.
\end{multline*}

Next, observe that almost surely, for any $c>0$,
\beq \label{eq:ch1}
|\hat S_{n,\tau}(\beta) - \tilde S_{n,\tau}(\beta)| \leq \sup_{|v|\geq c} |\Kc(v/h_n) - I\{v\geq 0\}| + (1+\sup_v |\Kc(v)|)\frac{1}{n}\sum_{i=1}^n
I\{|\beta^T W_i| \leq ch_n\}.
\eeq
Moreover,
\[
I\{|\beta^T W_i| \leq ch_n\} = I\{\beta^T W_i \leq ch_n\} - I\{\beta^T W_i > -ch_n\},
\]
and the classes of functions $\{w \mapsto I\{w^T\beta \leq c\}|\beta \in \R^{d+1}, c\in \R\}$, $\{w \mapsto I\{w^T\beta > c\}|\beta \in \R^{d+1}, c\in \R\}$ are VC-subgraph by Lemma 2.6.15 and Lemma 2.6.18 in \cite{vaarwell1996}. In combination with Theorem 2.6.7 and Theorem 2.4.3 from the same reference this implies
\[
\sup_{\beta\in \R^{d+1}}\sup_{c\in\R} \Big|\frac{1}{n}\sum_{i=1}^n I\{|\beta^T W_i| \leq ch_n\} - P(|\beta^T W_i| \leq ch_n)\Big| \to 0 \quad a.s.
\]
Setting $c = c_n = h_n^{-1/2}$ in the bound for $|\hat S_{n,\tau}(\beta) - \tilde S_{n,\tau}(\beta)|$ we see that the first term in \eqref{eq:ch1}, which is independent of $\beta$, converges to zero by assumption (K1). Moreover, by assumption (F1) we have for $\beta = (1,b^T)^T$
\bean
\sup_{b\in\R^d}P( |Z_i+b^T X_i| \leq c_nh_n) &=& \sup_{b\in\R^d}\int F_{Z|X}(-b^Tx + h_n^{1/2}|x) - F_{Z|X}(-b^Tx-h_n^{1/2}|x) dP_X(x)
\\
&=& o(1)
\eean
almost surely. A similar result holds for $\beta = (-1,b^T)^T$. Combining all the results so far we thus see that 
\[
k_n:= \sup_{b\in\R^d,s=\pm 1}\Big|\hat S_{n,\tau}((s,b^T)^T)-S_\tau((s,b^T)^T)\Big| = o(1) \quad a.s.
\]
Finally, observe that any $\beta := (s_0,b^T)^T \in \{-1,1\}\times B$ fixed
\[
\hat S_{n,\tau}(\beta) - \hat S_{n,\tau}(\bar \beta_\tau) \leq S_{\tau}(\beta) -  S_{\tau}(\bar \beta_\tau) + 2 k_n \leq -d(\|\beta-\bar \beta_\tau\|) + 2k_n.
\]
Since by definition $\hat S_{n,\tau}((\hat s_\tau,\hat b_\tau^T)^T) \geq \hat S_{n,\tau}((s,b^T)^T)$ for all $s \in \{-1,1\}, b \in B$, it follows that $k_n < d(\eps)/2$ implies $\sup_{\tau \in T} \|(\hat s_\tau,\hat b_\tau) - (s_0,\bar b_\tau)\|\leq \eps$. This completes the proof. \hfill $\Box$

\bigskip

\textbf{Proof of Theorem \ref{theo:betahat}}
Define 
\bea
\tilde S_n(s,b,\tau) &:=& \frac{1}{n}\sum_{i=1}^n (Y_i - (1-\tau))\Kc\Big(\frac{X_i^Tb + sZ_i}{h_n} \Big).
\eea
First, by uniform consistency of $\hat \beta_\tau$ and given the fact that $s_0 \in \{-1,1\}$ we see that with probability tending to one $\hat s_\tau = s_0$ for all $\tau\in T$. Thus we see that with probability tending to one $\hat b_\tau$ will satisfy
\[
\hat b_\tau = \mbox{argmax}_{b\in B} \hat S_n(s_0,b,\tau).
\]
Moreover, uniform consistency of $\hat b_\tau$ implies that with probability tending to one it will satisfy $\sup_{\tau\in T}\|\hat b_\tau - \bar b_\tau\|\leq\eta$, and continuous differentiability of $b \mapsto \tilde S_n(s,b,\tau)$ for $b$ with $\|b-\bar b_\tau\|\leq \eta$ implies that, with probability tending to one,
\[
\tilde T_n(s_0,\hat b_\tau,\tau) = 0 \quad \forall \tau \in T.
\]
A Taylor expansion now yields that with probability tending to one 
\beq \label{eq0}
\tilde T_n(s_0,\bar b_\tau,\tau) + \tilde Q_n(s_0,b^*_\tau,\tau)(\hat b_\tau - \bar b_\tau) = 0 \quad \forall \tau \in T
\eeq
where
\beq \label{eq:defqtild}
\tilde Q_n(s,b,\tau) := \frac{\partial \tilde T_n(s,b,\tau)}{\partial b} = \frac{1}{nh_n^2}\sum_{i=1}^n (Y_i - (1-\tau))X_iX_i^T K'\Big(\frac{X_i^Tb + sZ}{h_n} \Big)
\eeq
and $b^*_\tau = \xi_n(\tau)\bar b_\tau + (1-\xi_n(\tau))\hat b_\tau$ for some $\xi_n(\tau) \in [0,1]$. Define
\beq \label{eq:defq}
Q(s,b,\tau) := \int \frac{\partial}{\partial u}\Big((\tau - F_{Y^*|X,Z}(0|x,s(-x^Tb+u)))f_{Z|X}(s(-x^Tb+u)|x) \Big)\Big|_{u=0} xx^T dP_X(x).
\eeq
Rearranging \eqref{eq0} we obtain
\beq \label{eq:bah1}
Q(s_0,\bar b_\tau,\tau)(\hat b_\tau - \bar b_\tau) = - \tilde T_n(s_0,\bar b_\tau,\tau) - (\tilde Q_n(s_0,b^*_\tau,\tau) - Q(s_0,\bar b_\tau,\tau))(\hat b_\tau - \bar b_\tau).
\eeq
Since $\|\hat b_\tau - \bar b_\tau\|_\infty = o_P(1)$ uniformly over $\tau\in T$ and since the same holds for $\tilde T_n(s_0,\bar b_\tau,\tau)$ [see Lemma \ref{lem:t1}], there exists a $\gamma_n \to 0$ such that
\[
\sup_{\tau\in T}\Big\|- \tilde T_n(s_0,\bar b_\tau,\tau) - (\tilde Q_n(s_0,b^*_\tau,\tau) - Q(s_0,\bar b_\tau,\tau))(\hat b_\tau - \bar b_\tau) \Big\| = o_P(\gamma_n).
\]
By the conditions on $Q(s_0,\bar b_\tau,\tau)$ this implies $\sup_\tau \|\hat b_\tau - \bar b_\tau\| = o_P(\gamma_n)$. By Lemma \ref{lem:q1} this in turn implies [here, for matrix $A = (a_{ij})_{i,j}$ define $\|A\|_{max} := \max_{i,j}|a_{ij}|$]
\[
\sup_{\tau\in T} \Big\| \tilde Q_n(s_0,b^*_\tau,\tau) - Q(s_0,\bar b_\tau,\tau) \Big\|_{max} = O_P((nh_n^3)^{-1/2}\log n) + O(h_n+\alpha_n) + o_P(\gamma_n). 
\]
Finally, note that by \eqref{eq:qprime} we have $Q(s_0,\bar b_\tau,\tau) = Q_0(\tau)$.
Plugging this into (\ref{eq:bah1}) and repeating this argument [note that every application yields an improvement of the bound until $\gamma_n \sim \sup_\tau \|\tilde T_n(s_0,\bar b_\tau,\tau)\|$] yields the assertion (\ref{eq:bahres}).\\
For a proof of assertion (\ref{eq:betaweak}) fix $1 \leq j,j' \leq d$. Observe that
\begin{align*}
& \Cov\Big(\sqrt{nh_n}(\tilde T_n(s_0,\bar b_\tau,\tau))_j, \sqrt{nh_n}(\tilde T_n(s_0,\bar b_\kappa,\kappa))_{j'}\Big)
\\
= &\frac{1}{nh_n}\sum_{i,i' = 1}^n \Cov\Big((Y_i - (1-\tau))X_{ij} K\Big(\frac{X_i^T\bar b_\tau + s_0Z_i}{h_n}\Big),(Y_{i'} - (1-\kappa))X_{i'j'} K\Big(\frac{X_{i'}^T\bar b_\kappa + s_0Z_{i'}}{h_n} \Big)\Big)
\\
= & \frac{1}{h_n} \Cov\Big((Y_1 - (1-\tau))X_{1j} K\Big(\frac{X_1^T\bar b_\tau + s_0Z_1}{h_n}\Big),(Y_{1} - (1-\kappa))X_{1j'} K\Big(\frac{X_{1}^T\bar b_\kappa + s_0Z_{1}}{h_n} \Big)\Big).
\end{align*}
Now by the boundedness of the support of $\Xc$ and the assumptions on $K$
\[
\E\Big|(Y_1 - (1-\tau))X_{1j} K\Big(\frac{x^T\bar b_\tau + s_0z}{h_n}\Big) \Big| \leq C \int_\Xc \int_\R \Big|K\Big(\frac{X_1^T\bar b_\tau + s_0z}{h_n}\Big)\Big| f_{Z|X}(z|x) dz dP_X(x) = O(h_n).
\]
Thus it follows that for any $\delta > 0$
\begin{align}
& \Big|\Cov\Big(\sqrt{nh_n}(\tilde T_n(s_0,\bar b_\tau,\tau))_j, \sqrt{nh_n}(\tilde T_n(s_0,\bar b_\kappa,\kappa))_{j'}\Big)\Big| \nonumber
\\
\leq~ &\frac{C}{h_n}\int\int \Big|K\Big(\frac{x^T \bar b_\tau + s_0z}{h_n} \Big)K\Big(\frac{x^T\bar b_\kappa + s_0z}{h_n} \Big)\Big|f_{Z|X}(z|x)dzdP_X(x) \label{eq:boundcov}
 + O(h_n)
\\
\leq~ &\frac{C}{h_n} \int\Big(\sup_{|a|\geq \delta h_n^{-1}}|K(a)| + \|K\|_{\infty}I\{|w^T\beta_\tau|<|\beta_{\tau,1}|\delta\}\Big) \nonumber
\\
&\quad\quad\quad\times\Big(\sup_{|a|\geq \delta h_n^{-1}}|K(a)| + \|K\|_{\infty}I\{|w^T\beta_\kappa|<|\beta_{\kappa,1}|\delta\}\Big) dP_W(w)  + O(h_n) \nonumber
\end{align}
The assumptions on $K$ imply that for any $\delta>0$ we have $\sup_{|a|\geq \delta h_n^{-1}}|K(a)| = o(h_n^{k_0})$. Thus it remains to consider the integral
\[
\frac{1}{h_n}\int_\R I\{|W^T\beta_\kappa|<|\beta_{\kappa,1}|\delta\}I\{|W^T\beta_\tau|<|\beta_{\tau,1}|\delta\} dP_W(w) \leq \frac{1}{h_n} P\Big(|W^T(\beta_\tau - \beta_\kappa)| \leq 2\delta \max(|\beta_{\kappa,1}|,|\beta_{\tau,1}|)  \Big)
\] 
By assumption $w^T\beta_\tau$ is the conditional $\tau$-quantile of $Y^*$ given $W=w$. Hence, for any fixed $w \in \Wc$, $w^T\beta_\tau = F_{Y^*|W}^{-1}(\tau|w)$. Under assumption (D6) we have $|F_{Y^*|W}(y_1|w) - F_{Y^*|W}(y_2|w)| \leq |y_1-y_2| \sup_{y,w} f_{Y^*|W}(y|w) =: f_\infty |y_1-y_2|$ for any $w\in \Wc$. Substituting $y_1 = F_{Y^*|W}^{-1}(\tau|w) = w^T\beta_\tau, y_2 = F_{Y^*|W}^{-1}(\kappa|w) = w^T\beta_\kappa$ we find that $|w^T(\beta_\tau-\beta_\kappa)| \geq f_\infty^{-1} |\tau-\kappa|$ for any $w\in \Wc$. Hence
\[
P\Big(|W^T(\beta_\tau - \beta_\kappa)| \leq 2\delta \max(|\beta_{\kappa,1}|,|\beta_{\tau,1}|)  \Big) = 0
\]
provided that $2 \max(|\beta_{\kappa,1}|,|\beta_{\tau,1}|)\delta < f_\infty^{-1} |\tau - \kappa|/2$. Since $\delta > 0$ can be chosen to be arbitrarily small we obtain that   
\[
\Cov\Big(\sqrt{nh_n}(\tilde T_n(s_0,\bar b_\tau,\tau))_j, \sqrt{nh_n}(\tilde T_n(s_0,\bar b_\kappa,\kappa))_{j'}\Big) = o(1)
\]
for any $1 \leq j,j'\leq d$ and $\tau \neq \kappa$. The rest of the proof follows by standard arguments and is omitted.
\hfill $\Box$\\

\newpage

\begin{lemma}\label{lem:q1}
Under assumptions (A), (M), (K1), (K2), (B), (D2), (D3), (D5) we have [$\alpha_n$ was defined in assumption (K2)]
\bea
&&\sup_{\tau \in T}\sup_{\|b - \bar b_\tau\| \leq \eta} \Big\|\E[\tilde Q_n(s_0,b,\tau)] -  Q(s_0,b,\tau)\Big\|_{max} = O(h_n+\alpha_n), \label{eq:qew}
\\
&& \sup_{\tau \in T}\sup_{\|b - \bar b_\tau\| \leq \eta} \Big\|\tilde Q_n(s_0,b,\tau) - \E[\tilde Q_n(s_0,b,\tau)]\Big\|_{max} = O_P((nh_n^3)^{-1/2}\log n), \label{eq:qpr}
\eea
where $\tilde Q_n(s,b,\tau)$ was defined in (\ref{eq:defqtild}) and $Q(s_0,b,\tau)$ in \eqref{eq:defq}.
Moreover, we have for any $a_n \to 0$ 
\beq \label{eq:diffq}
\sup_{\tau\in T}\sup_{\|b-\bar b_\tau\|\leq a_n} \Big\|Q(s_0,\bar b_\tau,\tau) - Q(s_0,b,\tau) \Big\|_{max} \leq Ca_n
\eeq
for $a_n$ small enough and some universal constant $C$.
\end{lemma}
\textbf{Proof} 
We begin by considering assertion (\ref{eq:qew}). Observe that
\bean
&& \E[\tilde Q_n(s_0,b,\tau)] = \frac{1}{h_n^2}\int\int (\tau - F_{Y^*|X,Z}(0|x,z))f_{Z|X}(z|x) xx^T K'\Big(\frac{x^Tb + s_0z}{h_n} \Big)dzdP_X(x)
\\
&=& \frac{1}{h_n}\int\int \Big(\tau - F_{Y^*|X,Z}(0|x,s_0(vh_n-x^Tb))\Big) f_{Z|X}(s_0(vh_n-x^Tb)|x) xx^TK'(v)dv dP_X(x).
\eean
The assertion now follows from a Taylor expansion, the assumptions on $K$, and standard arguments similar to those given in \cite{horowitz2009}.
For a proof of (\ref{eq:qpr}) note that for $i,j = 1,...,d$ we have
\[
\sup_{\tau \in T}\sup_{\|b-\bar b_\tau\|\leq \eta} \Big|\Big(\tilde Q_n(s_0,b,\tau) - \E[\tilde Q_n(s_0,b,\tau)]\Big)_{i,j}\Big| \leq h_n^{-2}n^{-1/2}\sup_{f\in \Fc_n^{i,j}}|\Gb_n(f)|
\]
where $\Fc_n^{i,j}$ denotes the $n-$dependent class of functions
\[
\Fc_n^{i,j} := \Big\{f_{n,b,\tau}(x,y,z) =  K'\Big(\frac{x^T(\bar b_\tau + b) + s_0z}{h_n} \Big)(y - (1-\tau))x_ix_j \Big| b \in \R^d,\|b\|\leq \eta, \tau\in T\Big\}
\]
and 
\[
\Gb_n(f) := n^{-1/2}\sum_{i=1}^n (f(X_i,Y_i,Z_i) - \E[f(X_i,Y_i,Z_i)]).
\]
Now uniform H\"{o}lder continuity of $K'$, uniform H\"{o}lder continuity of $\tau \mapsto b_\tau$, and uniform boundedness of $x$ implies that for every sufficiently small $\delta>0$ we have for all $|\tau-\tau'|\leq\delta,\|b-b'\|\leq\delta$ for some $\gamma>0$
\[
\|f_{n,b,\tau}-f_{n,b',\tau'}\|_\infty \leq \frac{C}{h_n}(|\tau-\tau'|^\gamma + \|b-b'\|^\gamma)
\]
with $C$ denoting some constant independent of $n,\tau,\tau',b,b'$. This shows that for sufficiently small $\eps$ the $\|\cdot\|_\infty$-bracketing number [see \cite{vaarwell1996}, Chapter 2] of the class $\Fc_n^{i,j}$ is bounded by
\[
\Nc_{[~]}(\eps,\Fc_n^{i,j},\|\cdot\|_\infty) \leq \tilde C h_n^{-(d+1)/\gamma} \eps^{-(d+1)/\gamma}.
\] 
Next, observe that for any $\tau \in T, \|b\|\leq \eta $
\bean
\E[f_{n,b,\tau}^2(X,Y,Z)] &\leq& C \int\int_\R \Big| K'\Big(\frac{x^T(\bar b_\tau + b) + s_0z}{h_n} \Big)\Big|^2 dz dP_X(x) = C\int\int_\R |K'(zh_n^{-1})|^2dz dP_X(x)
\\
&=& h_n C \int_\R |K'(z)|^2dz. 
\eean
Combining this with Lemma \ref{lem:base} yields 
\[
\sup_{f\in \Fc_n^{i,j}}|\Gb_n(f)| = O_P(h_n^{1/2}\log n),
\]
and thus the proof of (\ref{eq:qpr}) is complete. Finally, assertion (\ref{eq:diffq}) follows by the smoothness properties of $F_{Y^*|X,Z}$ and $f_{Z|X}$. Thus the proof is complete. 
\hfill $\Box$

\begin{lemma}\label{lem:t1}
Under assumptions (A), (M), (K1)-(K3), (B), (D2), (D4), (D5) we have
\bea
&&\sup_{\tau \in T}\Big\|\E[\tilde T_n(s_0,\bar b_\tau,\tau)] -  T_n(s_0,\bar b_\tau,\tau)\Big\| = o(h_n^{k_0}), \label{eq:tew}
\\
&& \sup_{\tau \in T} \Big\|\tilde T_n(s_0,\bar b_\tau,\tau) - \E[\tilde T_n(s_0,\bar b_\tau,\tau)]\Big\| = O_P((nh_n)^{-1/2}\log n). \label{eq:tpr}
\eea

\end{lemma}
\textbf{Proof} The proof of (\ref{eq:tpr}) follows by arguments very similar to those used to establish (\ref{eq:qpr}) and is therefore omitted. For the proof of (\ref{eq:tew}), note that
\bean
&& \E[\tilde T_n(s_0,\bar b_\tau,\tau)] = \frac{1}{h_n}\int\int x(\tau - F_{Y^*|X,Z}(0|x,z)) K\Big(\frac{s_0 z + x^T\bar b_\tau}{h_n}\Big)f_{Z|X}(z|x) dz dP_X(x)
\\
&=& |s_0|\int\int x\Big(\tau - F_{Y^*|X,Z}(0|x,s_0(vh_n-x^T\bar b_\tau))\Big) f_{Z|X}(s_0(vh_n-x^T\bar b_\tau)|x) K(v)dv dP_X(x).
\\
&=& \int\int_{|vh_n|>\eta} x\Big(\tau - F_{Y^*|X,Z}(0|x,s_0(vh_n-x^T\bar b_\tau))\Big) f_{Z|X}(s_0(vh_n-x^T\bar b_\tau)|x) K(v)dv dP_X(x)
\\
&& + \int\int_{|vh_n|\leq\eta} x\Big(\tau - F_{Y^*|X,Z}(0|x,s_0(vh_n-x^T\bar b_\tau))\Big) f_{Z|X}(s_0(vh_n-x^T\bar b_\tau)|x) K(v)dv dP_X(x).
\eean
The order of the first integral is $o(h_n^{k_0})$ by the assumptions on $K$. The assertion now follows by a Taylor expansion of the function
\[
u \mapsto \Big(\tau - F_{Y^*|X,Z}(0|x,s_0(u-x^T\bar b_\tau))\Big) f_{Z|X}(s_0(u-x^T\bar b_\tau)|x),
\]
which holds for $|u|\leq \eta$ the assumptions on $K$ and standard arguments. \hfill $\Box$\\
\\

\newpage

\subsection{Proofs for Section~\ref{sec:prob} }

\textbf{Proof of Theorem \ref{theo:pw}} We will prove the result by applying Theorem~\ref{theo:rear} with $Q = \Wc_0$, $g_w(\tau) := w^T \bar\beta_\tau I\{\tau\in (a_w,b_w)\} - M I\{\tau<a_w\} + MI\{\tau>b_w\}$, $g_{n,w}(\tau) := w^T \hat\beta_\tau I\{\tau\in (a_w,b_w)\} - M I\{\tau<a_w\} + MI\{\tau>b_w\}$ where $M>0$ is a fixed constant. Note that this definition ensures that $\hat p_w(a_w,b_w) = \Psi_{g_{n,w}}(0), p_w = \Psi_{g_{w}}(0)$. Since $g_w(\tau) = 0$ is equivalent $w^T \beta_\tau = 0$, $g_w$ has a unique zero in $u_{0,w} = \tau_w$. Next, observe that $\tau \mapsto g_w(\tau)$ is continuously differentiable on $T^\delta$. Its derivative is given by
\[
\frac{\partial}{\partial \tau} w^T \bar \beta_{\tau} = \frac{s_0}{\beta_{\tau,1}} \Big[\frac{\partial}{\partial \tau} w^T \beta_{\tau}\Big] + w^T\beta_\tau \frac{\partial}{\partial \tau}\frac{s_0}{\beta_{\tau,1}} = \frac{s_0}{\beta_{\tau,1}f_{Y^*|W}(q_\tau(w)|w)}  + w^T\beta_\tau \frac{\partial}{\partial \tau}\frac{s_0}{\beta_{\tau,1}}
\]   
where we used the fact that
\[
\frac{\partial}{\partial \tau} w^T \beta_{\tau} = \frac{\partial}{\partial \tau} F_{Y^*|W}^{-1}(\tau|w) = \frac{1}{f_{Y^*|W}(q_\tau(w)|w)}.
\]
This shows that the derivative of $\tau \mapsto g_w(\tau)$ is uniformly H\"older continuous with exponent and constant uniform in $w \in \Wc_0$. Additionally, the derivative of $\tau \mapsto g_w(\tau)$ in the point $\tau_w$ is given by $(|\beta_{\tau_w,1}|f_{Y^*|W}(0|w))^{-1}$, and thus is bounded away from zero uniformly over $w \in \Wc_0$. Next, we show that for all $\eps >0$ we have
\beq \label{eq:cont1}
\inf_{w\in\Wc_0} \inf_{|\tau-\tau_w| > \eps, \tau \in T^\delta} |g_w(\tau)-g_w(\tau_w)| > 0.
\eeq
This can be proved by contradiction. To this end, observe that for $w \in \Wc_0$ fixed we have $\inf_{|\tau-\tau_w| > \eps, \tau \in T^\delta} |g_w(\tau)-g_w(\tau_w)| > 0$ for all $\eps>0$, this follows by compactness of $T^\delta$ and since $\tau\mapsto g_w(\tau)$ has a unique zero on $T^\delta$. Moreover, under the assumptions made one can prove that $w \mapsto \tau_w$ and $(w,\tau) \mapsto g_w(\tau)$ are continuous. Now the proof by contradiction follows by the same arguments as given in the last paragraph of the proof that (D1) follows from (A) and (S). The details are omitted for the sake of brevity. To apply Theorem~\ref{theo:rear}, it remains to establish that \eqref{eq:cpsi1}-\eqref{eq:cpsi3} hold. Here, \eqref{eq:cpsi1} follows from Lemma~\ref{lem:betaequi} which we prove next while \eqref{eq:cpsi2}, \eqref{eq:cpsi3} follow from the definition of $g_w, g_{n,w}$, Theorem~\ref{lem:unifcons} and Lemma~\ref{lem:t1}. This yields \eqref{eq:bahpw}. The rest of the proofs follows by standard arguments and is omitted. $\hfill \Box$

\begin{lemma}\label{lem:betaequi}
Under the assumptions of Theorem \ref{theo:pw} we have for any $r_n = o(h_n)$
\[
\sup_{w \in \Wc_0}\sup_{|\tau-\tau_w|\leq r_n} \|\hat\beta_{\tau_w} - \hat \beta_{\tau}-(\bar \beta_{\tau_w}-\bar \beta_{\tau})\| = O_P(n^{-1/2}h_n^{-3/2}r_n\log n) + o_P(h_n^{k_0}) + O_P(\kappa_n)
\]
where $\kappa_n$ is defined in Theorem \ref{theo:betahat}.
\end{lemma}

\textbf{Proof} From Theorem \ref{theo:betahat} we know that $\hat s_\tau = s_0$ for all $\tau \in T^\delta$ with probability tending to one, and thus it suffices to find a bound for $\|\hat b_\tau - \bar b_\tau - (\hat b_{\tau_w} - \bar b_{\tau_w})\|$. Here we have for any $\tau,\tau_w \in T^\delta$
\bean
&&\hat b_\tau - \bar b_\tau - (\hat b_{\tau_w} - \bar b_{\tau_w})
\\
&=& - Q_0(\tau)^{-1}\tilde T_n(s_0,\bar b_\tau,\tau) + Q_0(\tau_w)^{-1}\tilde T_n(s_0,\bar b_{\tau_w},{\tau_w}) - Q_0(\tau)^{-1}R_n(\tau) + Q_0(\tau_w)^{-1}R_n(\tau_w)
\\
&=&  Q_0(\tau_w)^{-1}\Big(\tilde T_n(s_0,\bar b_{\tau_w},{\tau_w})-\tilde T_n(s_0,\bar b_\tau,\tau)\Big) 
\\
&&+ \Big( Q_0(\tau_w)^{-1}-  Q_0(\tau)^{-1}\Big)\tilde T_n(s_0,\bar b_\tau,\tau) - Q_0(\tau)^{-1}R_n(\tau) + Q_0(\tau_w)^{-1}R_n(\tau_w)
\eean
with $R_n$ defined in \eqref{eq:bahres}. Combining conditions (D3) and (T) with the results from Theorem \ref{theo:betahat} and Lemma \ref{lem:t1} we see that the term in the last line is of order $O_P(r_n(nh_n)^{-1/2}\log n  + \kappa_n) + o_P(h_n^{k_0})$. For the first term, note that
\[
\sup_{w \in \Wc_0}\sup_{|\tau-\tau_w|\leq r_n} \Big\|\tilde T_n(s_0,\bar b_{\tau_w},{\tau_w})-\tilde T_n(s_0,\bar b_\tau,\tau)\Big\| \leq  D_{1,n} + D_{2,n}
\]
where
\bean
D_{1,n} &:=& \Big(\sum_{j=1,...,d}\sup_{\|f-g\|_2\leq C h_n^{-1/2}r_n, f,g\in \Fc_{n,j}} h_n^{-1}n^{-1/2}|\Gb_{n}(f)-\Gb_{n}(g)|^2\Big)^{1/2},
\\
D_{2,n} &:=& \sup_{w \in \Wc_0}\sup_{|\tau-\tau_w|\leq r_n}\Big\|\E[\tilde T_n(s_0,\bar b_{\tau_w},{\tau_w})-\tilde T_n(s_0,\bar b_\tau,\tau)] \Big\|, 
\\
\Gb_n(f) &:=& \frac{1}{n}\sum_{i=1}^n (f(Z_i,X_i,Y_i) - \E[f(Z_i,X_i,Y_i)]),
\eean
and the classes of functions $\Fc_{n,j}$ are given by
\[
\Fc_{n,j} = \Big\{(z,x,y) \mapsto K\Big(\frac{x^T(\bar b_\tau+b) + s_0z}{h_n} \Big)x_j(y - (1-\tau))\Big| b\in \R^d, \|b\|\leq\eta,\tau \in T^\delta \Big\}.
\]
In order to see that the representation for $D_{1,n}$ is true, note that 
\bean
&&\tilde T_n(s_0,\bar b_{\tau_w},{\tau_w})-\tilde T_n(s_0,\bar b_\tau,\tau)
\\
&=& \frac{1}{nh_n}\sum_{i=1}^n \Big(K\Big(\frac{X_i^T\bar b_\tau + s_0Z_i}{h_n} \Big)(Y_i - (1-\tau)) - K\Big(\frac{X_i^T\bar b_{\tau_w} + s_0Z_i}{h_n} \Big)(Y_i - (1-\tau_w))\Big)X_i.
\eean
In particular we have for $|\tau-\tau_w|\leq r_n$ and $n$ large enough
\bean
&&\Big|K\Big(\frac{X_i^T\bar b_\tau + s_0Z_i}{h_n} \Big)(Y_i - (1-\tau)) - K\Big(\frac{X_i^T\bar b_{\tau_w} + s_0Z_i}{h_n} \Big)(Y_i - (1-\tau_w))\Big|
\\
&\leq& \sup_{|v|\leq 1}\Big|K'\Big(v + \frac{X_i^T\bar b_\tau + s_0Z_i}{h_n}\Big)\Big|\frac{|X_i^T\bar b_\tau - X_i^T\bar b_{\tau_w}|}{h_n}
 + \sup_v |K(v)||\tau-\tau_w|.
\eean
This shows that for sufficiently small $\eps$ the $\|\cdot\|_\infty$-bracketing number [see \cite{vaarwell1996}, Chapter 2] of the class $\Fc_{n,j}$ is bounded by
\[
\Nc_{[~]}(\eps,\Fc_{n,j},\|\cdot\|_\infty) \leq C h_n^{-(d+1)/\gamma} \eps^{-(d+1)/\gamma}.
\] 
Moreover, the above bound implies that for $|\tau-\tau_w|\leq r_n$ we have for $n$ large enough
\bean
\sup_{w \in \Wc_0}\sup_{|\tau-\tau_w|\leq r_n}\Big\|K\Big(\frac{X_i^T\bar b_\tau + s_0Z_i}{h_n} \Big)(Y_i - (1-\tau)) - K\Big(\frac{X_i^T\bar b_{\tau_w} + s_0Z_i}{h_n} \Big)(Y_i - (1-\tau_w))\Big\|_{2,P}
\leq Ch_n^{-1/2}r_n 
\eean 
where $\|f\|_{2,P} := (\E[f^2(Z_i,X_i,Y_i)])^{1/2}$. Applying Lemma \ref{lem:base} to the classes of functions $\{f-g|f,g\in \Fc_{n,j},\|f-g\|_{2,P}\leq Ch_n^{-1/2}r_n\}$ thus shows that $D_{1,n} = O_P(n^{-1/2}h_n^{-3/2}r_n\log n).$\\
Next, consider $D_{2,n}$. By the results in Lemma \ref{lem:t1} we have
\begin{multline*}
|D_{2,n}| \leq h_n^{k_0}C_k \sup_{x \in \Xc}\sup_{w \in \Wc_0}\sup_{|\tau-\tau_w|\leq r_n}\Big|\frac{\partial^k}{\partial u^k}\Big(\Big((\tau - F_{Y^*|X,Z}(0|x,s_0(-x^T\bar b_{\tau}+u)))f_{Z|X}(s_0(-x^T\bar b_\tau+u)|x)  
\\
- (\tau_w - F_{Y^*|X,Z}(0|x,s_0(-x^T\bar b_{\tau_w}+u)))f_{Z|X}(s_0(-x^T\bar b_{\tau_w}+u)|x)\Big)\Big|_{u=0}\Big| + o(h_n^{k_0}).
\end{multline*}
From condition (D4) we see that the left-hand side in the above expression is of order $o(h_n^{k_0})$. Thus the proof is complete. \hfill $\Box$

\subsection{Proofs for Section~\ref{sec:rearr} }

\textbf{Proof of Theorem \ref{theo:rear}} 
The statement (\ref{eq:psi1}) is a direct consequence of the condition on the collection of functions $g_q$ and assumption (\ref{eq:cpsi2}).\\
The main technical ingredient for the remaining proof is the expansion in Lemma \ref{lem:rear}. By the assumptions on the collection $g_q$ and on $\chi$ there exists a constant $C_\chi < \infty$ such that 
\[
\inf_q \inf_{|u-u_{0,q}| > C_\chi R_n}|g_q(u)| > 2R_n.
\]
Together with conditions \eqref{eq:cpsi2}, \eqref{eq:cpsi3} this implies that for sufficiently large $n$ the signs of $g_q$ and $g_{n,q}$ coincide of the set $\{u: \delta > |u-u_{0,q}| > C_\chi R_n \}$ for all $q \in Q$. Together with \eqref{eq:psi1} this implies for sufficiently large $n$ and any $C \geq C_\chi$
\[
\Psi_{g_q}(0) - \Psi_{g_{q,n}}(0) = \int_{u_{0,q}-CR_n}^{u_{0,q}+CR_n} I\{g_{q}(u)\leq 0\} - I\{g_{n,q}(u)\leq 0\} du.
\] 
Next observe that $\sup_q |g_{n,q}(u_{0,q})| \leq R_n$. Let $C_0 := \max\{C_\chi,g_{min}^{-1}\}$. Then for all $q \in Q$ we have $g_q'(u_{0,q})C_0 R_n \geq R_n \geq |g_{n,q}(u_{0,q})|$. Hence Lemma~\ref{lem:rear} implies for any $q \in Q$
\[
\Big|\int_{u_{0,q}-C_0R_n}^{u_{0,q}+C_0R_n}  I\{g_q(u)\leq 0\} du - \int_{u_{0,q}-C_0R_n}^{u_{0,q}+C_0R_n}  I\{g_{n,q}(u)\leq 0\} du - \frac{g_{n,q}(u_{0,q})}{g_q'(u_{0,q})} \Big| \leq \frac{2\xi(C_0R_n) + 4 C_)R_n \chi(C_0R_n)}{g_q'(u_{0,q})}.
\]
Take a supremum over $q$ first on the right and then on the left to complete the proof. \hfill $\Box$ 

\newpage

\begin{lemma}\label{lem:rear}
Consider functions $g,h:[0,1]\to\R$ and assume that for some $u_{0}\in(0,1)$ we have $g(u_{0})=0$. Additionally, assume that $g$ is continuously differentiable in a neighborhood $U_\delta(u_{0})\subset (0,1)$ and that $g'(u_0) > 0$. Define 
\[
\xi(\eps) := \sup_{|u-u_0|\leq\eps} |h(u_0)-h(u) -(g(u_0)-g(u))|, \quad \chi(\eps) := \sup_{|u-u_0|\leq \eps}|g'(u) - g'(u_0)|.
\] 
Then for any $\eps<\delta$ with $|h(u_0)| \leq g'(u_0)\eps$
\[
\Big|\int_{u_0-\eps}^{u_0+\eps}  I\{g(u)\leq 0\} du - \int_{u_0-\eps}^{u_0+\eps}  I\{h(u)\leq 0\} du - \frac{h(u_0)}{g'(u_0)} \Big| \leq \frac{2\xi(\eps) + 4 \eps \chi(\eps)}{g'(u_0)}.
\]
\end{lemma}
\textbf{Proof}. Rewrite
\[
I\{h(u)\leq 0\} = I\{h(u_0) - (g(u_0)-g(u)) - (h(u_0)-h(u) -(g(u_0)-g(u))) \leq 0\}
\]
and observe that by the properties of $g$ we have 
\[
\sup_{|u-u_0|\leq\eps} |g(u)-g(u_0) - g'(u_0)(u-u_0)| \leq \eps \chi(\eps).
\]
Thus the indicators $I\{h(u_0) + g'(u_0)(u-u_0)\leq 0,|u-u_0|\leq\eps\}$ and $I\{h(u_0) - (g(u_0)-g(u)) - (h(u_0)-h(u) -(g(u_0)-g(u))) \leq 0,|u-u_0|\leq\eps\}$ can only take different values on a set with Lebesgue measure at most $2(\xi(\eps) + \eps \chi(\eps))/g'(u_0)$. This implies
\[
\Big|\int_{u_0-\eps}^{u_0+\eps} I\{h(u)\leq 0\} du - \int_{u_0-\eps}^{u_0+\eps} I\{h(u_0) + g'(u_0)(u-u_0)\leq 0\} du\Big| \leq \frac{2(\xi(\eps) + \eps \chi(\eps))}{g'(u_0)}.
\]
Recalling that $g(u_0)=0$, similar arguments yield the bound
\[
\Big|\int_{u_0-\eps}^{u_0+\eps} I\{g(u)\leq 0\} du - \int_{u_0-\eps}^{u_0+\eps} I\{g'(u_0)(u-u_0)\leq 0\} du\Big| \leq \frac{2\eps \chi(\eps)}{g'(u_0)}.
\]
Finally, a simple computation shows that for $|h(u_0)| \leq g'(u_0)\eps$
\[
\int_{u_0-\eps}^{u_0+\eps} I\{g'(u_0)(u-u_0)\leq 0\} du - \int_{u_0-\eps}^{u_0+\eps} I\{h(u_0) + g'(u_0)(u-u_0)\leq 0\} du = \frac{h(u_0)}{g'(u_0)}.
\]
Thus the proof is complete. \hfill $\Box$\\
\\

\begin{appendix}
\section{Technical details}

\begin{lemma}\label{lem:base}\  Assume that the classes of measurable functions $\ef_n$ consist of uniformly bounded functions (by a constant not depending on $n$). If additionally 
\[
N_{[\, ]}(\ef_n,\eps,L^2(P))\leq Cn^a\eps^{-a}
\]
for every $\eps\leq \delta_n$ and constants $C,a>0$ not depending on $n$, then we have for any $\delta_n \sim n^{-b}$ with $b<1/2$
\[
\sqrt{n} \sup_{f\in \ef_n, \|f\|_{P,2} \leq \delta_n} \Big(\int f dP_n - \int f dP\Big) = O_P^*\Big( \delta_n (|\log \delta_n| + \log n)\Big).
\]
Here, the $^*$ denotes outer probability if the supremum is not measurable [see Chapter 1 in \cite{vaarwell1996} for a more detailed discussion].
\end{lemma}
\textbf{Proof.}
Start by observing that the uniform boundedness of elements of $\ef_n$ by $D$ implies that $F \equiv D$ is a measurable envelope function with $L_2$-norm $D$. Note that for $\eta_n$ sufficiently small
\bean
\gamma(\eta_n) &:=& \eta_n D/\sqrt{1 + \log N_{[ ]}(\eta_n D,\ef_n,L_2(P))} \geq D\eta_n /\sqrt{1 + \log C + a\log n -a \log(D\eta_n)}
\\
&\geq& D\tilde C \eta_n/\sqrt{|\log \eta_n| + \log n}
\eean
for some finite constant $\tilde C$ depending only on $a,C,D$. Thus the bound in Theorem 2.14.2 in \cite{vaarwell1996} yields for $\delta_n$ sufficiently small [with $\E^*$ denoting outer expectation]
\bean
\E^*\Big[\sup_{f\in\ef_n} \Big|\int f d\alpha_n\Big|\Big] &\leq& DJ_{[~ ]}(\delta_n,\ef_n,L_2(P)) + \sqrt{n}\int F(u)I\{F(u)>\sqrt{n}\gamma(\delta_n)\}P(du)
\\
&\leq& D C_1\int_0^{\delta_n} |\log \eps| + \log n d\eps + D\sqrt{n}I\Big\{D>\frac{D\tilde C \sqrt{n}\delta_n}{|\log\delta_n|+ \log n} \Big\}
\\
&\leq& D C_2\delta_n(|\log \delta_n| + \log n) + D\sqrt{n}I\Big\{1>\frac{\tilde C \sqrt{n}\delta_n}{|\log\delta_n|+ \log n} \Big\}
\eean
where $\alpha_n := \sqrt{n}(P_n - P)$, $P_n$ denotes the empirical measure, and $C_1,C_2$ are some finite constants. Here, the second inequality follows by a straightforward calculation and the first inequality is due to the fact that for $\delta_n$ sufficiently small by definition
\bean
J_{[ ]}(\delta_n,\ef_n,L_2(P)) = \int_0^{\delta_n} \sqrt{1 + \log N_{[ ]}(\eps D,\ef_n,L_2(P))}d\eps \leq C_1 \int_0^{\delta_n} |\log \eps| + \log n d\eps.
\eean
Now under the assumption on $\delta_n$, the indicator $I\{|\log\delta_n|+ \log n >\tilde C \sqrt{n}\delta_n\}$ will be zero for $n$ large enough and thus the proof is complete. \hfill $\Box$

\newpage

\section{Additional simulation results} \label{sec:addsim}

This section contains additional tables and figures for Design 2 and Design 3 in Section~\ref{sec:sim}.

\begin{figure}
\includegraphics[width = 3.5in]{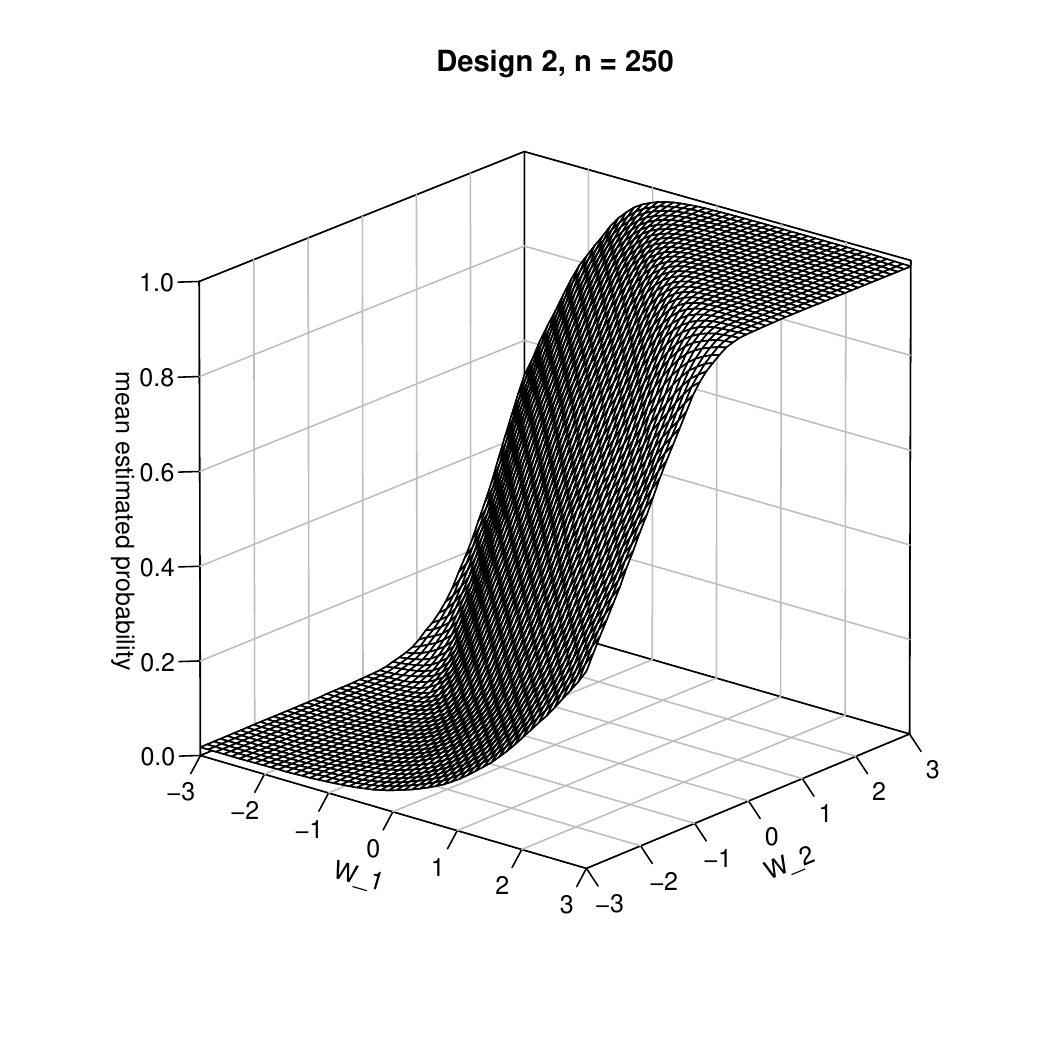}
\includegraphics[width = 3.5in]{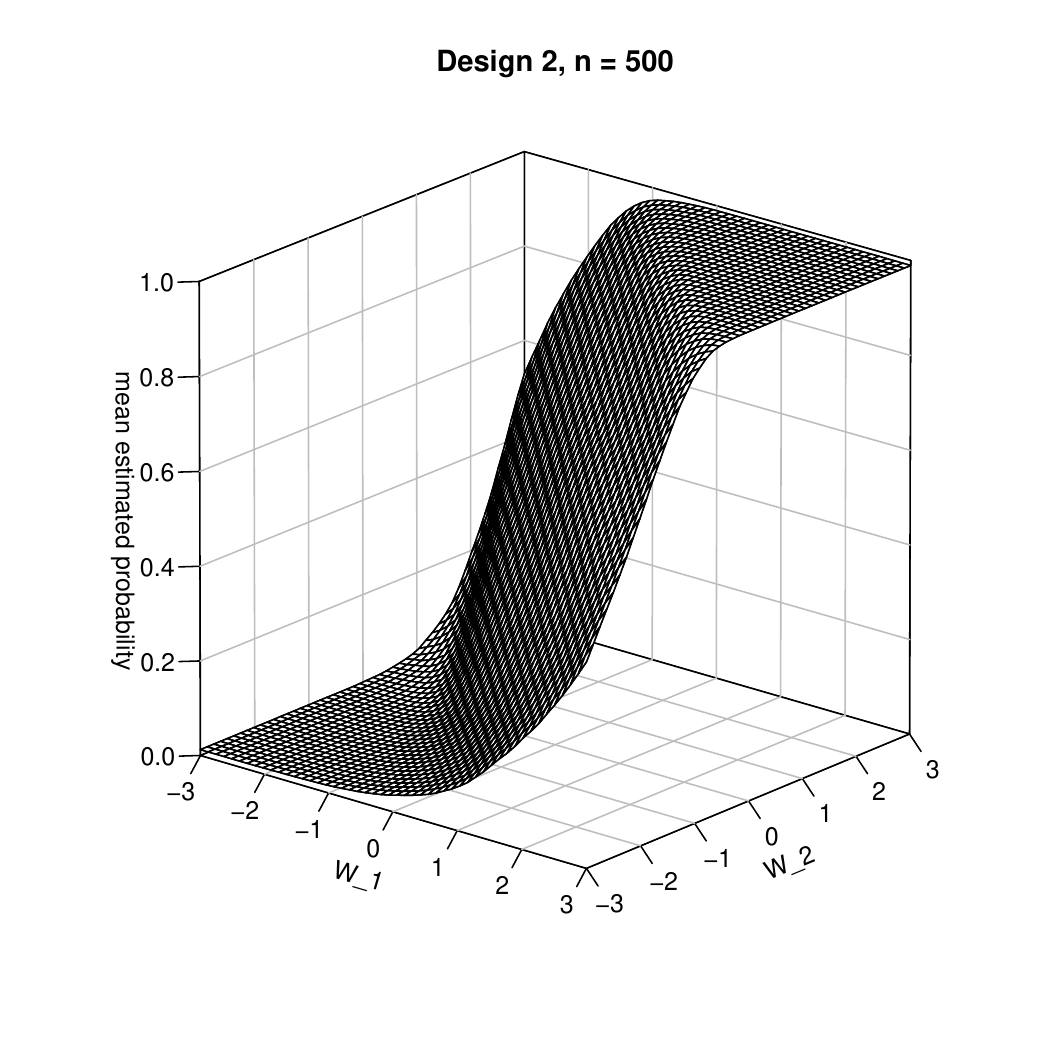}
\includegraphics[width = 3.5in]{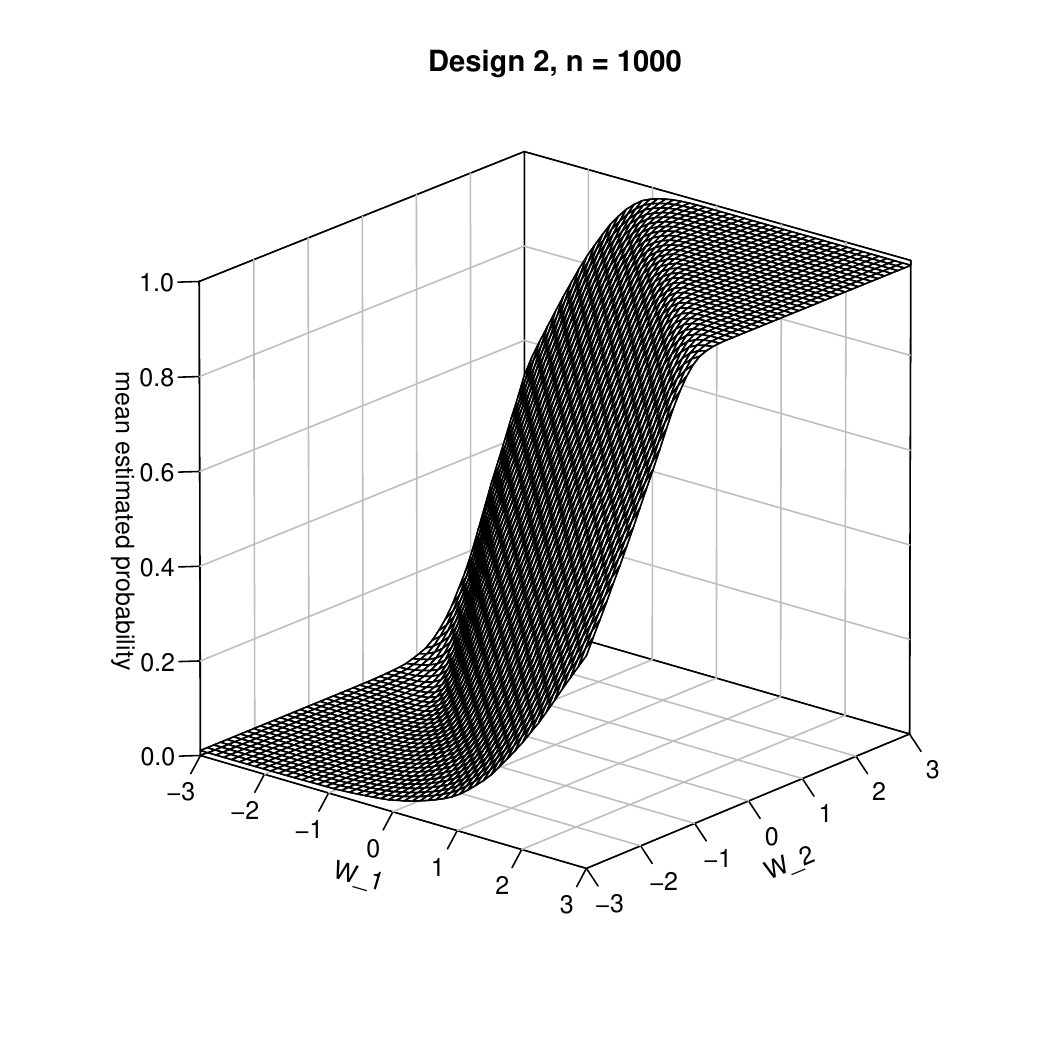}
\includegraphics[width = 3.5in]{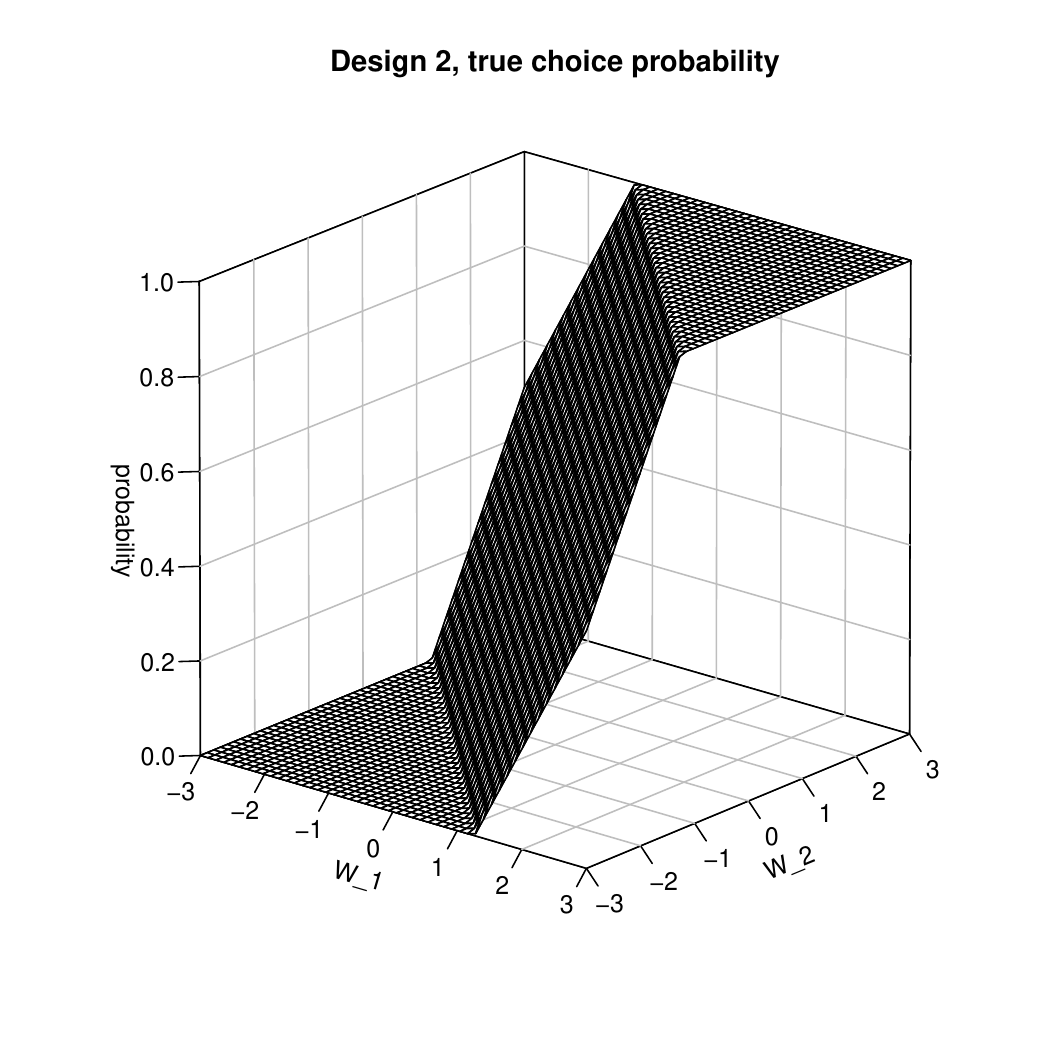} 
\caption{Average (over 500 simulations) estimated (first 3 figures) and true (bottom right) choice probabilities as functions of covariates in Design 2 } \label{fig2}
\end{figure} 

\begin{figure}
\includegraphics[width = 3.5in]{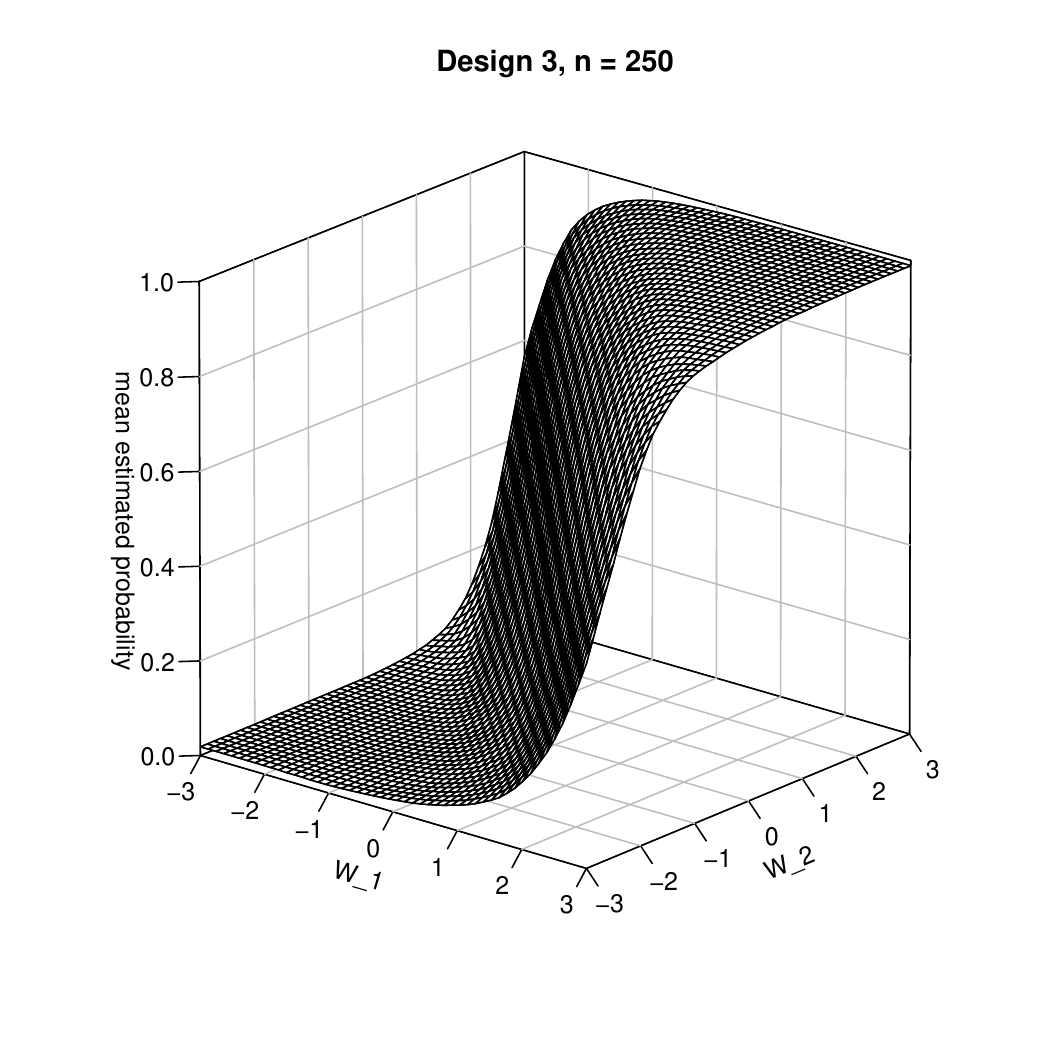}
\includegraphics[width = 3.5in]{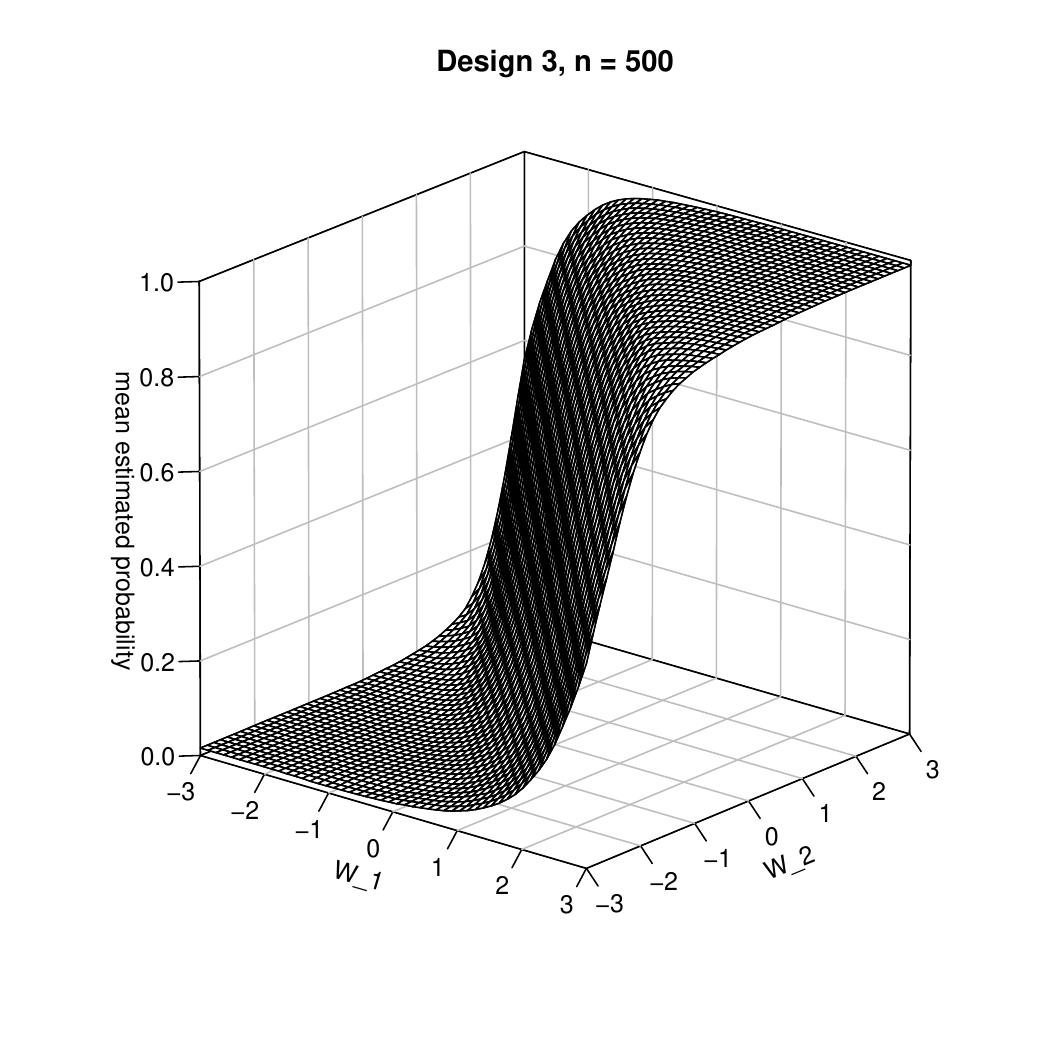}
\includegraphics[width = 3.5in]{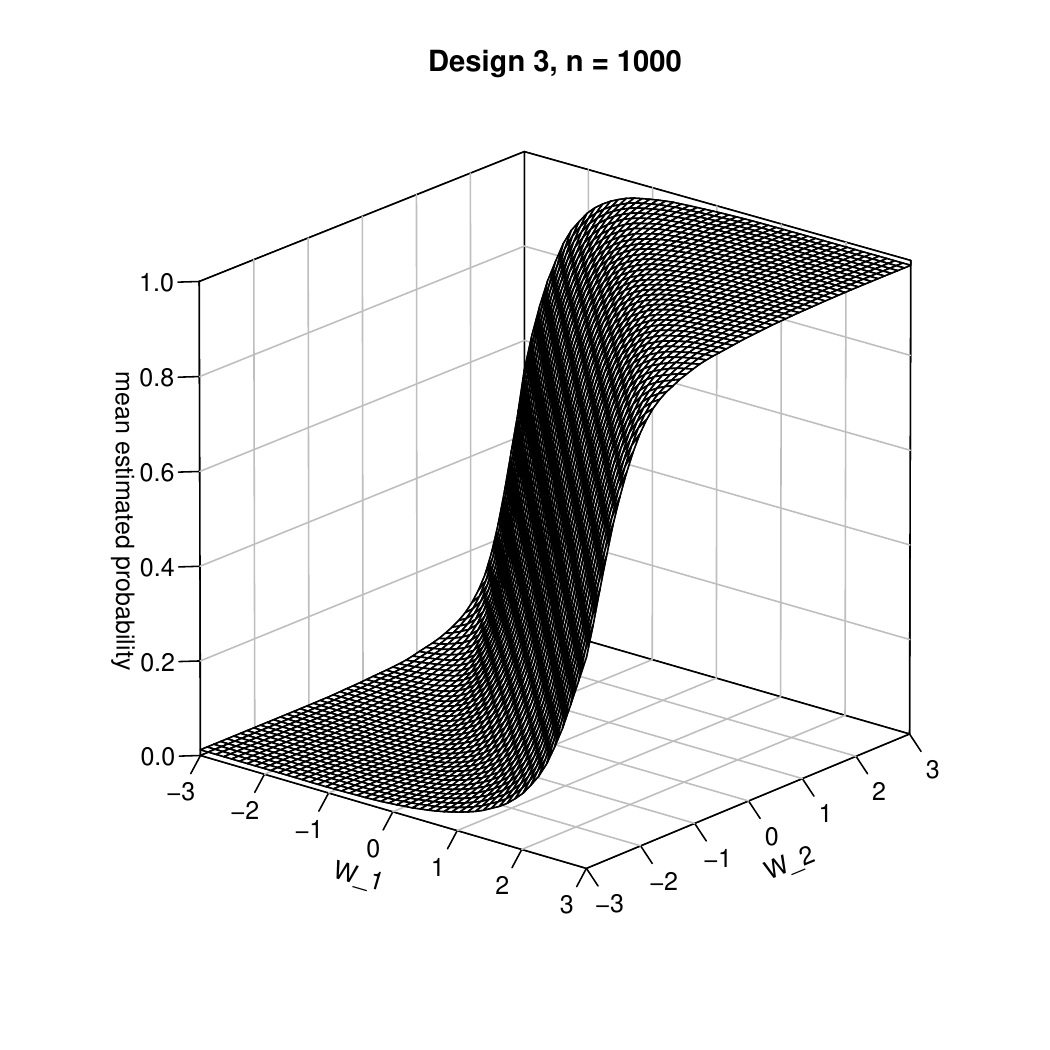}
\includegraphics[width = 3.5in]{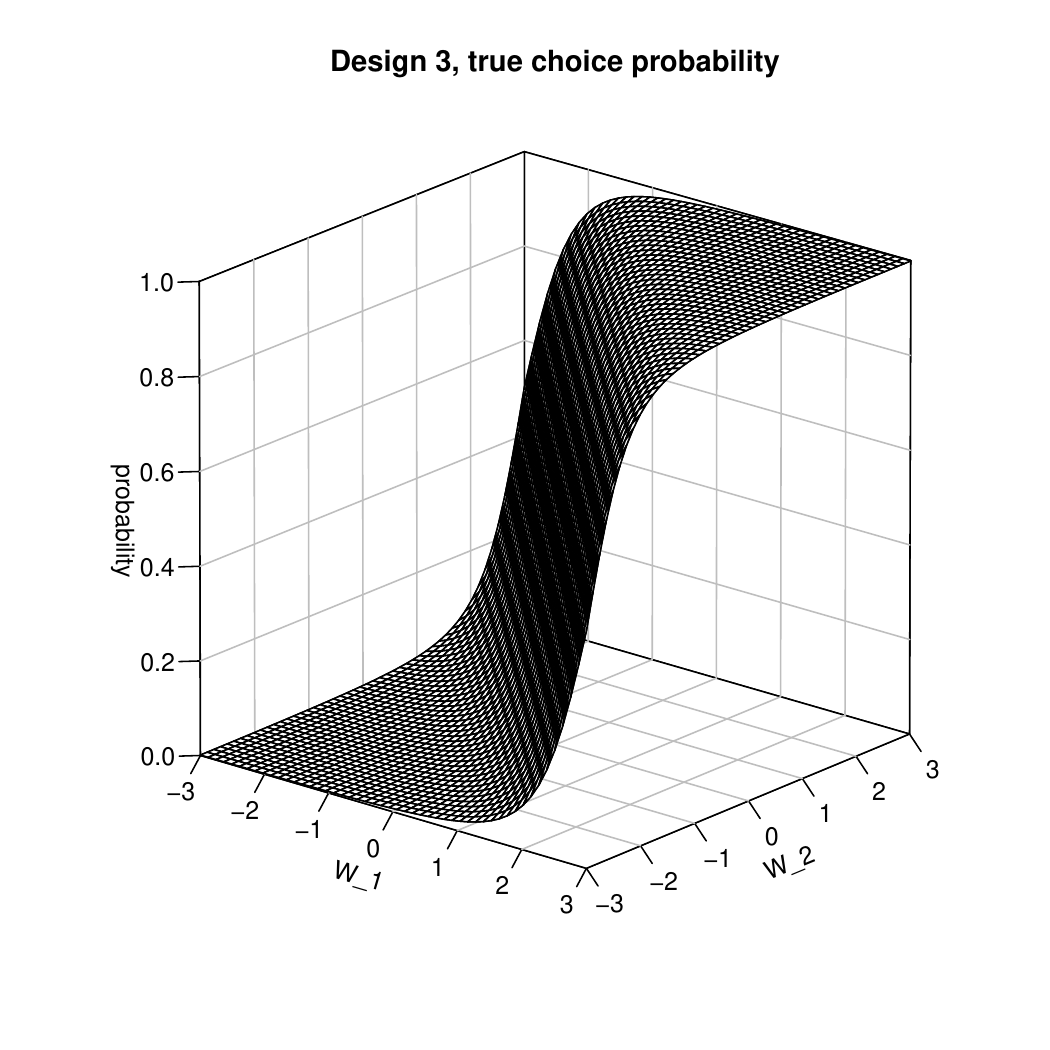} 
\caption{Average (over 500 simulations) estimated (first 3 figures) and true (bottom right) choice probabilities as functions of covariates in Design 3 } \label{fig3}
\end{figure}

\begin{table}[h] \small
\begin{tabular}{cc||c|c|c|c|c|c|c|c|c}
  & & 1-a~~1-b &1-a~~1-b&1-a~~1-b&1-a~~1-b&1-a~~1-b&1-a~~1-b&1-a~~1-b&1-a~~1-b&1-a~~1-b\\
  & & 0.01~0.99&0.15~0.99&0.25~0.99&0.01~0.85&0.15~0.85&0.25~0.85&0.01~0.75&0.15~0.75&0.25~0.75\\
	\hline
	$p_w$&  &\multicolumn{9}{c}{n = 250 }
	\\
	\hline
0.21& RMSE & 9.95& 8.2& 8.12& 9.91& 8.15& 8.09& 9.87& 8.11& 8.06\\
&bias& 0.7& 2.44& 6.51& 0.63& 2.4& 6.5& 0.57& 2.36& 6.48\\
\hline
0.5& RMSE &9.26& 9.23& 9.08& 9.23& 9.19& 9.02& 9.12& 9.06& 8.88\\
&bias& -0.05& 0.11& 0.3& -0.21& -0.05& 0.15& -0.39& -0.24& -0.04\\
\hline
0.79& RMSE &6.08& 6.05& 6.01& 5.79& 5.76& 5.72& 6.04& 6.02& 6.01\\
&bias& -1.4& -1.33& -1.27& -1.88& -1.85& -1.81& -5.34& -5.33& -5.33\\
\hline
&  &\multicolumn{9}{c}{n = 500 }
	\\
\hline
0.21& RMSE &7.84& 6.58& 6.59& 7.8& 6.54& 6.57& 7.77& 6.5& 6.55\\
&bias& 0.82& 1.93& 5.76& 0.75& 1.89& 5.75& 0.68& 1.86& 5.74\\
\hline
0.5& RMSE &6.83& 6.81& 6.8& 6.8& 6.77& 6.75& 6.76& 6.71& 6.67\\
&bias& 0.12& 0.28& 0.43& -0.05& 0.12& 0.28& -0.21& -0.05& 0.11\\
\hline
0.79&RMSE & 4.75& 4.71& 4.68& 4.69& 4.66& 4.62& 5.21& 5.2& 5.19\\
&bias& -1.5& -1.43& -1.36& -1.83& -1.8& -1.76& -4.82& -4.82& -4.81\\
\hline
&  &\multicolumn{9}{c}{n = 1000 }
	\\
\hline
0.21&RMSE & 5.67& 5.2& 5.53& 5.64& 5.17& 5.51& 5.61& 5.14& 5.5\\
&bias& 1.18& 1.73& 5.04& 1.11& 1.69& 5.04& 1.04& 1.66& 5.03\\
\hline
0.5&RMSE & 4.7& 4.7& 4.7& 4.68& 4.66& 4.66& 4.66& 4.63& 4.61\\
&bias& 0.34& 0.5& 0.65& 0.17& 0.33& 0.49& 0.02& 0.17& 0.33\\
\hline
0.79&RMSE & 3.47& 3.44& 3.41& 3.52& 3.49& 3.47& 4.46& 4.45& 4.45\\
&bias& -1.15& -1.08& -1.02& -1.43& -1.39& -1.36& -4.28& -4.28& -4.28\\
\hline
\end{tabular} 
\caption{\small Estimation quality of conditional choice probabilities for different values of $a,b$. Root mean squared error (RMSE) and bias are multiplied by $100$. Data are generated from Design 2.}	\label{tab3}
\end{table}

\begin{table}[h] \small
\begin{tabular}{cc||c|c|c|c|c|c|c|c|c}
  & & 1-a~~1-b &1-a~~1-b&1-a~~1-b&1-a~~1-b&1-a~~1-b&1-a~~1-b&1-a~~1-b&1-a~~1-b&1-a~~1-b\\
  & & 0.01~0.99&0.15~0.99&0.25~0.99&0.01~0.85&0.15~0.85&0.25~0.85&0.01~0.75&0.15~0.75&0.25~0.75\\
	\hline
	$p_w$&  &\multicolumn{9}{c}{n = 250 }
	\\
	\hline
  0.09 & RMSE &7.53& 8.13& 16.05& 7.5& 8.12& 16.05& 7.47& 8.1& 16.05 \\
 &bias& 1.49& 7.46& 16.03& 1.46& 7.46& 16.03& 1.43& 7.45& 16.03 \\
\hline
   0.5& RMSE &11.04& 11.02& 10.83& 10.99& 10.96& 10.75& 10.6& 10.54& 10.31 \\
 &bias& 0.96& 1.12& 1.34& 0.8& 0.96& 1.18& 0.51& 0.66& 0.88 \\
\hline
   0.91&RMSE &5.13& 5.1& 5.07& 6.78& 6.77& 6.77& 15.92& 15.92& 15.92 \\
 &bias&-2.32& -2.28& -2.25& -6.56& -6.56& -6.56& -15.92& -15.92& -15.92 \\
\hline
&  &\multicolumn{9}{c}{n = 500 }
	\\
\hline
 0.09&RMSE & 5.7& 6.96& 15.92& 5.68& 6.96& 15.92& 5.65& 6.95& 15.92\\
&bias& 1.48& 6.68& 15.92& 1.45& 6.67& 15.92& 1.42& 6.67& 15.92\\
\hline
 0.5&RMSE & 7.91& 7.9& 7.85& 7.87& 7.85& 7.79& 7.8& 7.76& 7.68\\
&bias& 0.66& 0.82& 0.98& 0.49& 0.65& 0.82& 0.33& 0.48& 0.65\\
\hline
 0.91&RMSE & 3.96& 3.93& 3.9& 6.24& 6.24& 6.24& 15.92& 15.92& 15.92\\
&bias& -1.95& -1.92& -1.89& -6.15& -6.15& -6.15& -15.92& -15.92& -15.92\\
\hline
&  &\multicolumn{9}{c}{n = 1000 }
	\\
\hline
0.09&RMSE & 4.47& 6.39& 15.92& 4.44& 6.39& 15.92& 4.42& 6.39& 15.92\\
&bias& 1.41& 6.26& 15.92& 1.38& 6.26& 15.92& 1.35& 6.26& 15.92\\
\hline
0.5&RMSE & 5.51& 5.51& 5.51& 5.47& 5.46& 5.46& 5.45& 5.42& 5.4\\
&bias& 0.54& 0.71& 0.86& 0.38& 0.54& 0.7& 0.23& 0.38& 0.53\\
\hline
0.91&RMSE & 3.01& 2.98& 2.96& 6& 6& 6& 15.92& 15.92& 15.92\\
&bias& -1.48& -1.45& -1.42& -5.99& -5.99& -5.99& -15.92& -15.92& -15.92\\
\hline
\end{tabular} 
\caption{\small Estimation quality of conditional choice probabilities for different values of $a,b$. Root mean squared error (RMSE) and bias are multiplied by $100$. Data are generated from Design 3.}	\label{tab4}
\end{table}

\end{appendix}

\end{document}